\documentclass{jpp}
\usepackage{graphicx}

\usepackage[utf8]{inputenc}
\usepackage[T1]{fontenc}
\usepackage{amsmath}
\usepackage{amssymb}
\usepackage{xcolor}

\newcommand{\bfm}[1]{\mbox{\boldmath$#1$}}

\shorttitle{The maximum-$J$ property in quasi-isodynamic stellarators}
\shortauthor{E. Rodríguez, P. Helander, and A. G. Goodman}

\title{The maximum-$J$ property in quasi-isodynamic stellarators}

\author{E. Rodríguez, P. Helander, and A. G. Goodman}

\affiliation{Max Planck Institute for Plasma Physics, 17491 Greifswald, Germany}

\begin{document}

\maketitle

\begin{abstract}
Some stellarators tend to benefit from favourable average magnetic curvature for trapped particles when the plasma pressure is sufficiently high. This so-called maximum-$J$-property has several positive implications such as good fast-particle confinement, magnetohydrodynamic stability, and suppression of certain trapped-particle instabilities. This property cannot be attained in quasisymmetric stellarators, in which deeply trapped particles experience average bad curvature and therefore precess in the diamagnetic direction close to the magnetic axis. However, quasi-isodynamic stellarators offer greater flexibility and allow the average curvature to be favourable and the precession to be reversed. We find that it is possible to design such stellarators so that the maximum-$J$ condition is satisfied for the great majority of all particles, even when the plasma pressure vanishes. The qualitative properties of such a stellarator field can be derived analytically by examining the most deeply and the most shallowly trapped particles, although some small fraction of the latter will inevitably not behave as desired. However, through numerical optimisation, we construct a vacuum field in which 99.6\% of all trapped particles satisfy the maximum-$J$ condition. 

\end{abstract}

\section{Introduction}\label{sec:intro}

A basic requirement for magnetic fusion devices is the ability to confine collisionless particle orbits, a property which not all stellarators possess. Although the existence of nested magnetic surface is sufficient to confine circulating (untrapped) particle orbits to lowest order in the smallness of the gyroradius \citep{helander2014theory}, particles that are trapped in local minima along the magnetic field tend to drift out of the plasma unless the field is carefully tailored to avoid this phenomenon.

If we write the magnetic field in Clebsch form \citep{d2012flux}, $\mathbf{B} = \nabla \psi \times \nabla \alpha$, then the coordinates $(\psi,\alpha)$ are constant along $\bf B$ and can thus be used to label field lines. In a field tracing out toroidal magnetic surfaces, we take $\psi$ to be the magnetic flux enclosed by such a surface divided by $2 \pi$, and $\alpha \in [0,2\pi]$ thus labels the different field lines on each surface. In Boozer coordinates $\alpha = \theta - \iota \varphi$, where $\theta$ denotes the poloidal angle, $\varphi$ the toroidal angle, and $\iota(\psi)$ the rotational transform \citep{boozer1981plasma}. 

As a trapped particle drifts across the field, the second adiabatic invariant \citep{hastie1967adiabatic}
    $$ \mathcal{J}_\parallel = \int m v_\| \mathrm{d}\ell $$
is conserved to high accuracy if the gyroradius is small. Here, $m$ denotes the mass of the particle in question, $v_\| = v \sqrt{1-\lambda B}$ its speed along the magnetic field line (if the electric field is sufficiently small), and $\ell$ the arc length in this direction. The ratio of the magnetic moment to the kinetic energy is denoted by $\lambda = v_\perp^2/(v^2 B)$, where $B$ is the field strength, and the integration is carried out between two consecutive bounce points where $v_\|$ vanishes. $\mathcal{J}_\|$ is thus a function of the variables $(\psi,\alpha,v,\lambda)$. 

The net cross-field magnetic drift of a magnetically trapped particle of charge $q$ travelling between two consecutive bounce points is given by the Hamiltonian equations \citep{kadomtsev1967plasma,helander2014theory}
\begin{subequations}
    \begin{equation} \Delta \psi = \frac{1}{q} \frac{\partial \mathcal{J}_\|}{\partial \alpha}, 
    \label{Delta psi}
    \end{equation}
    \begin{equation} \Delta \alpha = - \frac{1}{q} \frac{\partial \mathcal{J}_\|}{\partial \psi}. 
    \label{Delta alpha}
    \end{equation}
\end{subequations}
In particular, the particle remains in the vicinity of one flux surface if $\p \mathcal{J}_\|/\p \alpha = 0$. Magnetic fields in which this property holds for all orbits are called omnigenous \citep{Hall,Cary1997}. The direction and magnitude with which particles precess within the surface is governed by $\p \mathcal{J}_\parallel/\partial \psi$. It is convenient to define a precession frequency $\omega_\alpha$, by dividing $\Delta\alpha$ by the bounce time $\Delta t=\int\mathrm{d}\ell/v_\parallel$. Fields in which this quantity is negative for electrons (that is, $q \omega_\alpha > 0$) and thus opposed to the diamagnetic drift ($q\omega_\star\sim p'(\psi)/n<0$), so-called {\em maximum-$J$-configurations}, have long been known to possess favourable stability properties for trapped-particle modes \citep{rosenbluth1968,proll2012resilience,helander2013collisionless}. 

The maximum-$J$ property is additionally correlated with, but not identical to \citep{helander2014theory}, the existence of a magnetic well \citep{greene1997}, which is beneficial for MHD-stability. This circumstance has to do with its relation to magnetic curvature. If the magnetic curvature vector is decomposed into components in the direction of $\nabla \psi$ and $\nabla \alpha$,
    $$ \bfm{\kappa} = {\bf b} \cdot \nabla {\bf b} = \kappa_\psi \nabla \psi + \kappa_\alpha \nabla \alpha, $$
then a positive value of $ \kappa_\psi$ is referred to as favourable (or ``good'') curvature. Since the guiding-centre curvature drift is in the direction ${\bf b} \times \bfm{\kappa}$ and
    $$ \kappa_\psi = \frac{({\bf b} \times \bfm{\kappa}) \cdot \nabla \alpha}{B}, $$
good curvature is clearly correlated with the maximum-$J$ property. Indeed, the latter can be interpreted as the condition that a certain average of the normal curvature should be favourable for all trapped-particle orbits. 

The trapped particles in an omnigenous stellarator can precess in the poloidal, toroidal or helical directions, and if the precession is poloidal, the field is called {\em quasi-isodynamic} (QI) \citep{Helander_2009,Nührenberg_2010}. This feature is determined by the topology of the contours of constant $|\mathbf{B}|$. The aim of the present paper is to explore conditions under which a QI field may possess the maximum-$J$-property. It has long been known that this can be the case at finite plasma $\beta$ (thermal pressure divided by magnetic pressure), very recently achieved at low values \citep{sanchez2023quasi,goodman2024} and, as we shall see, it is also possible (to a very good approximation) when $\beta$ vanishes. This is in stark contrast to quasisymmetric magnetic fields (including tokamaks), where the maximum-$J$-property is unattainable, at least in a region near the magnetic axis. If the maximum-$J$ condition is satisfied for some, but not all, particle orbits, then ${\partial \mathcal{J}_\|}/{\partial \psi} $ must vanish for certain orbits and these will usually undergo large radial excursions, resuling in  \textit{super-banana} \citep{velasco2021model} or \textit{banana-drift convective} \citep{paul2022energetic} losses. It is clearly preferable that all orbits satisfy the maximum-$J$ condtion. 

Quasi-isodynamic stellarators are sometimes referred to as a system of linked-mirrors \citep{Boozer_1998}. Following this analogy, the maximum-$J$-condition can be seen to be closely related to the so-called minimum-$B$ property in magnetic mirrors, which was proposed in the early 1960's \citep{Taylor_1963}. To see why, consider particles trapped in the vicinity of a local minimum, $B_{\rm min}(\psi)$. Since
    \begin{equation} \frac{\partial \mathcal{J}_\parallel}{\partial \psi} = - \frac{mv}{2} \int \left( \frac{\partial B}{\partial \psi} \right)_{\alpha,l} 
     \frac{\lambda dl}{\sqrt{1-\lambda B}}, 
    \label{dJdpsi}
    \end{equation}
it follows that the maximum-$J$-property requires
    $$ \left( \frac{\partial B}{\partial \psi} \right)_{\alpha,l} > 0$$
at the field minimum. In other words, $B_{\rm min}$ increases with radius and the magnetic field strength assumes a global minimum on the magnetic axis of a maximum-$J$ stellarator \citep{helander2014theory}. In the vicinity of this minimum, the surfaces of constant field strength are ellipsoids aligned with the magnetic field.

In the following sections, we explore these ideas further and derive conditions for attaining the maximum-$J$-condition, focusing particularly on deeply trapped particles close to the magnetic axis. 

\section{Trapped-particle precession}

A derivation of the general expression for the precession frequency of trapped particles in an arbitrary stellarator is presented in Appendix A. The result is expressed as an integral along the magnetic field between two consecutive bounce points, which in Boozer coordinates becomes
\begin{equation}
       \omega_\alpha = \frac{mv^2}{q} \int \left[  \left(\frac{\partial B}{\partial \psi}\right)_{\alpha, \varphi} 
    \frac{1-\lambda B/2}{B^2 \sqrt{1-\lambda B}} 
    +\frac{\mu_0 p'(\psi)}{\langle B^2 \rangle}
    \frac{\sqrt{1-\lambda B}}{B} \right] d \varphi   
      \bigg\slash \int \frac{d \varphi}{B \sqrt{1-\lambda B}}.
    \label{omegaa-simple}
    \end{equation}
For simplicity, we have assumed that the net toroidal plasma current enclosed by the magnetic surface under consideration vanishes, $I(\psi)=0$. 
\par
In Equation~(\ref{omegaa-simple}) there are two separate terms contributing to the precession: the radial derivative of $B=|\mathbf{B}|$ and the pressure gradient $p'$, which is negative for typical pressure profiles.\footnote{In the vacuum limit, the curvature drift and the grad-B drift are equal, and thus only the gradient of $B$ appears in the expression for $\omega_\alpha$, but if $\beta > 0$ this is generally not the case. The difference between the two drifts is then proportional to the pressure gradient, which is responsible for the additional pressure term in Eq.~(\ref{omegaa-simple}).} However, it would be wrong to conclude that a non-zero pressure gradient only affects the precession frequency through the explicit pressure term in Eq.~(\ref{omegaa-simple}). This na\"{i}ve interpretation would imply that a negative pressure gradient tends to make $q\omega_\alpha$ more negative and thus oppose the maximum-$J$ condition. However, the radial derivative of $|\mathbf{B}|$ depends on the pressure profile in an equilibrium field. If the central plasma $\beta$ is increased, the magnetic field strength usually drops in the centre of the plasma to hold the plasma in place, and $\partial_\psi B$ thus grows. The net result turns out, in practice, to make the maximum-$J$ condition much easier to attain at high plasma beta compared with the case of a vacuum field. It is thus important to address the question of how $\partial_\psi B$ depends on the pressure, which we shall consider later. 
\par
\begin{figure}
    \centering
    \includegraphics[width=\textwidth]{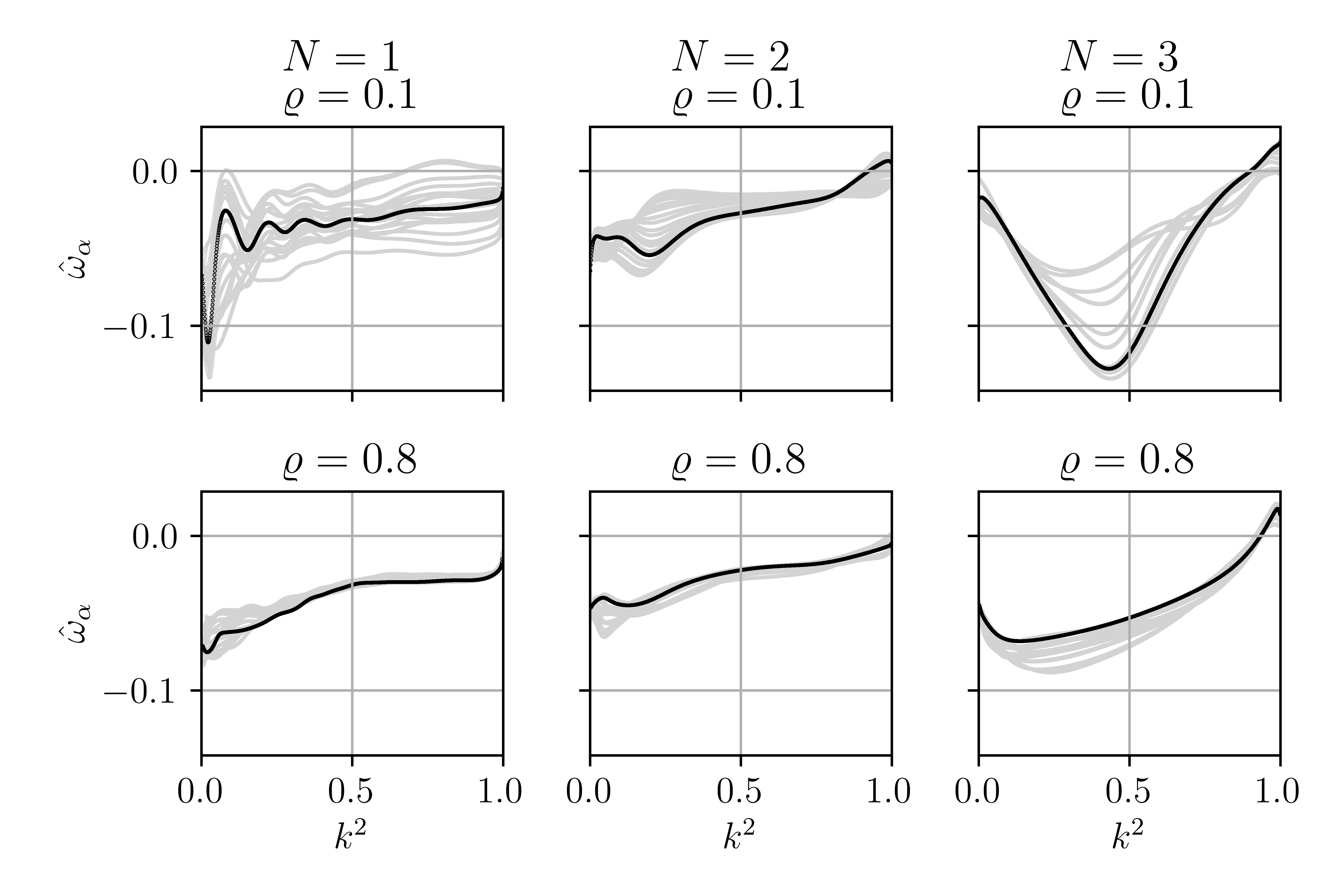}
    \caption{\textbf{Example of precession in highly quasi-isodynamic fields.} Normalised precession frequency $\hat{\omega}_\alpha$, Eq.~(\ref{omegaa-hat}), as a function of the trapping parameter $k^2$, Eq.~(\ref{k}), for three recently optimised quasi-isodynamic configurations with different number of field periods $N$ \citep{goodman2023constructing}, in which omnigeneity is attained to a high degree. For each configuration, $\hat{\omega}_\alpha$ is displayed at two values of the normalised radial flux coordinate $\varrho$ on a number of different field lines (gray curves), with the black curve representing $\alpha=0$. A positive value of $\hat{\omega}_\alpha$ corresponds to the maximum-$J$-property. Note that deeply trapped particles ($k^2 \rightarrow 0$) tend to behave worse than barely trapped ones, and the intermediate population can exhibit non-monotonic dependence on $k^2$.}
    \label{fig:qi-alan-examples-wa}
\end{figure}

In general, $\omega_\alpha$ is a function, for each trapping well, of the pitch-angle parameter $\lambda$, the flux-surface label $\psi$, and the field-line label $\alpha$. The dependence on $\alpha$ is present only when the field is not omnigeneous. Then the second adiabatic invariant depends on $\alpha$, and, as a result, the precession frequency of trapped particles varies as they drift from field line to field line. Numerical examples are shown in Figure~\ref{fig:qi-alan-examples-wa}, where the normalised precession frequency
  \begin{equation}
    \hat{\omega}_\alpha = \frac{4 q \psi_a \omega_\alpha}{m v^2}
        \label{omegaa-hat}
  \end{equation}
is displayed for three different numerically optimised QI fields \citep{goodman2023constructing}. Here $\psi_a$ denotes the value of $\psi$ at the plasma boundary and $\hat \omega_\alpha$ has been plotted as a function of the trapping parameter \citep{roach1995trapped},
  \begin{equation}
    k^2 = \frac{B_{\rm min}^{-1}-\lambda}{B_{\rm min}^{-1} - B_{\rm max}^{-1}}.
    \label{k}
  \end{equation}
As discussed below, the magnitude of $\hat{\omega}_\alpha$ can be interpreted as the precession frequency (for thermal particles) divided by the diamagnetic frequency (within a factor of order unity). 
The grey lines correspond to field lines on the same flux surface with different values of $\alpha$. The fact that these lines are different from each other is an indication of departure from exact omnigeneity. For the most part of the discussion to follow, we shall assume, unless otherwise stated, that the field is omnigeneous. We may then choose any field line to compute the bounce-average (\ref{dJdpsi}) of $\partial_\psi B$, or alternatively its Boozer form in (\ref{omegaa-simple}). In a vacuum field with $p'(\psi) = 0$, these integrals are weighted averages of $\partial_\psi B$, where the weight is positive for all trapped particles (i.e., all $\lambda$). It thus follows that the so-called minimum-$B$ condition, $\partial_\psi B>0$, is a \textit{sufficient} condition for maximum-$J$ behaviour. 

\par

\subsection{Necessary conditions: special points}
It is however not necessary to make $\partial_\psi B>0$ everywhere in order to attain the maximum-$J$ condition. Even if $\partial_\psi B < 0$ at certain positions along the field line, the average precession can still be favourable. The condition $\partial_\psi B>0$ must nevertheless be satisfied at all local minima and maxima of $|\mathbf{B}|$ along the field line. At these points, deeply and barely trapped particles spend nearly all their time, other points along the orbit not contributing much to the average. It is therefore \textit{necessary}, both at the bottom and top of every magnetic well, that $\partial_\psi B|_{\alpha,\varphi}>0$, even if $p'\neq0$. 
\par
This result may be formally derived as the appropriate limits of Eq.~(\ref{omegaa-simple}), but a simpler argument has already been given in Section \ref{sec:intro}: since the time required for a trapped particle to move from one bounce point to the next is equal to
    $$ \Delta t = \int \frac{dl}{v_\|} = \int \frac{dl}{v \sqrt{1-\lambda B}}, $$
the precession frequency for deeply trapped particles, for which $1-\lambda B \ll 1$,  becomes
    \begin{equation}
     \omega_\alpha = \frac{\Delta \alpha}{\Delta t} =
     \frac{mv^2}{2qB} \left(\frac{\partial B}{\partial \psi}\right)_{\alpha, l}.
    \label{omegaa0}
    \end{equation}
Since $\partial B/\partial \ell = 0$ at the point of minimum field strength, the derivative $\partial B / \partial \psi$ can equally well be computed at constant Boozer angles $(\theta,\varphi)$ for an omnigeneous stellarator. A similar argument holds for barely trapped particles. A more detailed discussion can be found in Appendix~\ref{app:precession}, including a proof that the explicit pressure term does not contribute, Eq.~(\ref{eqn:bot_well_walpha_p_vanish}).
\par
Thus, if the minimum-$B$-condition is satisfied at all local maxima and minima of $B$ along the field line, i.e. if 
\begin{equation}
    B'_\mathrm{max}(\psi)>0 \qquad \mbox{and} \qquad B'_\mathrm{min}(\psi)>0,     
    \label{eq:end points}
\end{equation}
then deeply and barely trapped particles are guaranteed to precess in the desired direction ($q \omega_\alpha > 0$). It would however be wrong to infer that all particles trapped at intermediate depths in the magnetic wells then precess in the same direction. Such a conclusion cannot be drawn by solely considering the radial gradient of $B$ at its extrema along the field. The conditions in Eq.~(\ref{eq:end points}) are \textit{necessary} for maximum-$J$, but not \textit{sufficient}. Enforcing them as part of an optimisation effort may be helpful, but does not guarantee the maximum-$J$ behaviour everywhere. 
\par
This inference is however possible under additional assumptions about $|\mathbf{B}|$. Obviously, if $\partial_\psi B$ were bounded from below by either of the local $|\mathbf{B}|$ gradients, then Eq.~(\ref{eq:end points}) would become a \textit{necessary} and \textit{sufficient} criterion. One scenario in which this holds true is when the shape of the magnetic well along field lines behaves `rigidly'. That is, in going from one surface to another, $|\mathbf{B}|=B_0(\psi)+\eta(\psi)f(\ell)$ along the field line. The case of a simple sinusoidal well in QI stellarators has recently been analysed and proposed as a potentially useful optimisation target \citep{velasco2023robust}.\footnote{The flat-mirror' criterion presented in \cite{velasco2023robust} guarantees Eq.~(\ref{eq:end points}) regardless of the model assumed for $|\mathbf{B}|$. In the  case $|B|=B_{00}+B_M\cos\varphi$ considered by the authors, this criterion implies maximum-$J$ behaviour of all trapped particles, but not in general.} In general, though, the behaviour of $\partial_\psi B$, and thus $\omega_\alpha$, can be rather complicated. Figure \ref{fig:qi-alan-examples-wa} shows a recently published example of omnigeneity-optimised QI stellarators, where the behaviour at $k=(0,1)$ (deeply and barely trapped particles respectively) are not always good indicators for intermediate particles. 
Nevertheless, given the simplicity and necessary nature of the condition in Eq.~(\ref{eq:end points}), we shall use it as a representative feature of the maximum-$J$-condition, and extend it to more global considerations.  

\subsection{Quasi-isodynamic fields}
The treatment so far has considered little information about the magnetic field other than properties that follow directly from the requirements of the MHD equilibrium and omnigeneity. In the interest of describing the particulars of QI fields, though, we introduce to the discussion features unique to this class of stellarators. To do so, and to further simplify the discussion, we focus our attention to the vicinity of the magnetic axis. 
\par
By expansion in the minor-radius coordinate $r = \sqrt{2 \psi / \bar B}$, where $\bar B$ is some reference magnetic-field strength, which we choose to be equal to that at the minimum, to second order the magnetic field magnitude may be written in the following form,
    $$ B = B_0(\varphi) + r B_1(\alpha,\varphi) + r^2 B_2(\alpha,\varphi), $$
where, as shown in Appendix A, the coefficients satisfy the following relation in an exactly QI stellarator-symmetric field
\citep{rodriguez2023higher}:
\begin{subequations}
\begin{gather}
     B_1(\alpha,\varphi) = - d(\varphi) \sin \alpha, \\
    B_2(\alpha,\varphi) = B_{20}(\varphi) - B_{2s}(\varphi) \sin 2 \alpha 
    - \frac{\partial}{\partial \varphi} \left( \frac{B_0^2 d^2}{4B_0'} \right) \cos 2\alpha. 
\end{gather}\label{eqn:QI_cond_nae}
\end{subequations}
Here $B_0(\varphi)$ and $B_{20}(\varphi)$ are even in $\varphi$ whereas $d(\varphi)$ and $B_{2s}(\varphi)$ are odd, if we choose the angle $\varphi$ to vanish at the point where $B_0(\varphi)$ attains its minimum. As a result, the first-order term $B_1$ vanishes at the minimum, and the precession frequency Eq.~(\ref{omegaa0}) for the most deeply trapped particles, being proportional to $\partial B/\partial \psi$, is determined by the second-order term $B_2$. It becomes
\begin{equation} 
    \omega_\alpha = \frac{mv^2}{qB_0^2} \left[ B_{20} - \left( \frac{B_0^2 d^2}{4B_0'} \right)' \right]+\omega_{\alpha}^\mathrm{non-QI},  \label{eqn:deep-QI-wa-nae}
\end{equation}
where everything is evaluated at the minimum, $\varphi = 0$, the primes denote a derivative with respect to the toroidal angle $\varphi$, and $\omega_{\alpha}^\mathrm{non-QI}$ represents the contribution from deviations from omnigeneity, which will be discussed later. To achieve $q\omega_\alpha>0$, the magnetic field must be carefully tailored in such a way that for deeply trapped particles the quantity in the square brackets of Eq.~(\ref{eqn:deep-QI-wa-nae}) is positive. As we shall see, this is eminently possible but requires careful consideration of the terms within the square brackets. 
\par
As shown explicitly in Appendix~\ref{sec:appB20Deep}, Eq.~(\ref{eqn:2nd-order-deviation-wa}), departures from the condition of omnigeneity lead to additional terms in the expression for precession, here denoted by $\omega_{\alpha}^\mathrm{non-QI}$. These contributions may be interpreted as non-intrinsic contributions due to departures from QI. Breaking omnigeneity at first order in $r$ leads to a contribution that scales as $\omega_{\alpha,-1}^\mathrm{non-QI}\propto\cos \alpha/r$, Eq.~(\ref{eqn:qi-1/r-wa}), while deviations at second-order lead to $\omega_{\alpha,0}^\mathrm{non-QI}\propto \cos2\alpha$, Eq.~(\ref{eqn:2nd-order-deviation-wa}). Because of the dependence on $\alpha$, there always exists a field line on which the non-omnigeneous contribution is $q\omega_\alpha^\mathrm{non-QI}<0$, i.e., detrimental to the maximum-$J$ condition. The deviations that arise at first order are particularly worrying near the magnetic axis due to the $1/r$ scaling. We explain the origin of this behaviour in the following subsection. From the forms above we may nevertheless conclude that omnigeneity is necessary for the maximum-$J$ property.

\subsection{Comparison with quasisymmetric fields}
\begin{figure}
        \centering
        \includegraphics[width=0.8\textwidth]{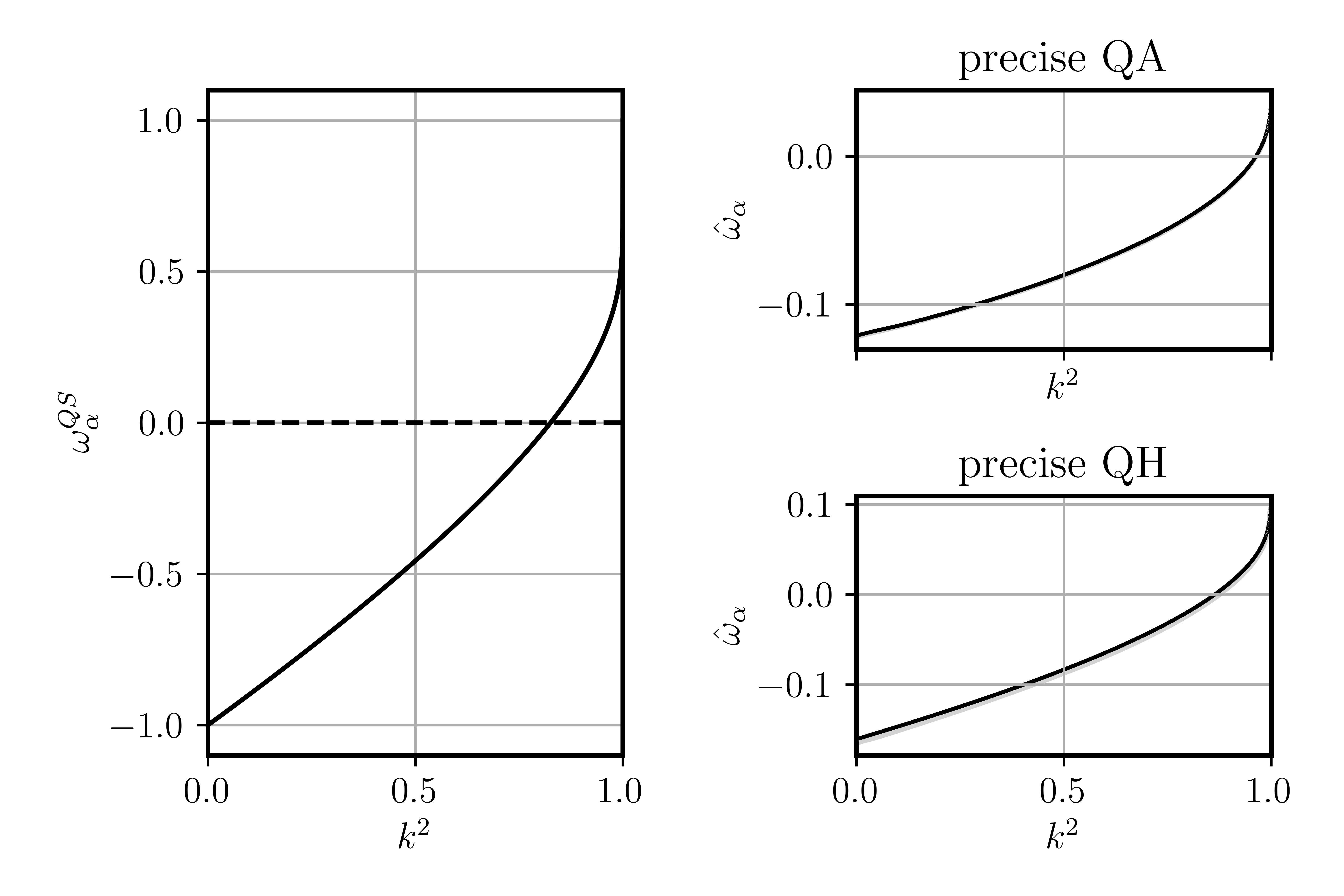}
        \caption{\textbf{Precession in zero-$\beta$ tokamaks and quasisymmetric stellarators.} The left plot shows the predicted dependence of the leading-order precession frequency in a tokamak or quasisymmetric stellarator as a function of the trapping parameter, $k$, Eq.~(\ref{eqn:QS_tok_precess}). The right plots show the precession frequency on the boundary of actual QS vacuum configurations \citep{landreman2022magnetic}, normalised as in Eq.~(\ref{omegaa-hat}) (see \cite{rodriguez2023precession} for more details and discussion). As in Figure~\ref{fig:qi-alan-examples-wa}, the plots on the right show the variation with field line label $\alpha$, which is difficult to discern due to the high degree of quasisymmetry in these configurations. }
        \label{fig:qs-wa}
    \end{figure}

Trapped-particle precession in a QI field is markedly different from that in quasisymmetric (QS) configurations, the other important class of optimised stellarators. Particle orbits in the latter are similar to those in tokamaks, and the maximum-$J$ property cannot be satisfied throughout the volume for all particle classes. In fact, close to the magnetic axis, the magnetic field strength in any quasisymmetric field (characterised by the helicity of the symmetry $N$) is  
\begin{equation}
    B(r,\chi=\theta-N\varphi) = B_0 (1 - r \eta \cos \chi) \label{eqn:QS_B_cond}
\end{equation}
and the precession frequency of trapped particles becomes \citep{rodriguez2023precession,kadomtsev1967plasma,helander2005collisional}
    \begin{equation}
        \omega_\alpha^\mathrm{QS} \approx - \frac{mv^2}{2q}\frac{\eta}{rB_0}\left(2\frac{E(k)}{K(k)}- 1 \right), \label{eqn:QS_tok_precess}
    \end{equation}
where $E$ and $K$ denote elliptic integrals of the first and second kind \citep[Sec.~19]{DLMF}. Their argument is defined by $k=\sin(\chi_b/2)$, where $\chi_b$ is the bounce point in the well, so that $k=0$ refers to deeply trapped particles and $k=1$ to the trapped-passing boundary. This definition matches Eq.~(\ref{k}) to leading order.  We present a plot of the behaviour in quasisymmetric configurations in Figure~\ref{fig:qs-wa}.
    \par
To make the comparison to the QI case more explicit, we turn to the general expression for the precession frequency close to the magnetic axis given in Eq.~(\ref{eqn:nae_w_alpha}) and consider the limit in which the magnetic-field strength only varies slightly along the magnetic axis. Then trapped particles have $1-\lambda B \ll 1$ and Eq.~(\ref{eqn:nae_w_alpha}) reduces to
  \begin{equation}
        \omega_{\alpha} = \frac{2mv^2}{q \bar{B}} \int \frac{f(\varphi)  d \varphi}{\sqrt{1 - \lambda B_0}}
        \bigg\slash \int \frac{d \varphi}{\sqrt{1 - \lambda B_0}}, \label{eqn:wa-QI-approx}
    \end{equation}
where
    \begin{equation} f(\varphi) = \frac{B_{20}(\varphi)}{B_0} - \frac{1}{B_0} \frac{d}{d\varphi} \left( \frac{B_0^2 d^2}{4 B_0'} \right). \label{eqn:fun_f_nae_qi}
    \end{equation}
The precession frequency is thus proportional to a bounce-average of the function $f(\varphi)$ between the two turning points. This expression reveals several important differences between the precession frequency in QI and QS stellarators (including tokamaks):

\begin{enumerate}
    \item In a QS field the particle precession becomes `infinite' ($\sim1/r$) as the magnetic axis is approached, whereas it remains finite in QI fields. This $1/r$ behaviour comes from a finite poloidal component of the curvature drift as the axis is approached. In a QI configuration, the curvature of the magnetic field vanishes at the minimum on the axis, eliminating this behaviour. In the QS case, deeply trapped particles reside in the bad-curvature region, and the normal curvature is even about the minimum of $B$ (see Figure~\ref{eqn:QS_B_cond}). This difference in behaviour follows directly from the difference in the topology of $|\mathbf{B}|$ contours, and its implications on the order-$r$ correction to $B_0$, Eqs.~(\ref{eqn:QI_cond_nae}) and (\ref{eqn:QS_B_cond}). Formally, this difference in parity explains the leading-order cancellation of the precession in QI fields and the elimination of the $1/r$ contribution. The cancellation ceases to be exact, though, whenever the field deviates from omnigeneity. In that event, there will always exist some $r_a$ such that, for $r<r_a$, the field ceases to satisfy the maximum-$J$ condition (see Fig.~\ref{fig:qi_analysis} for example). It is known that omnigeneity must be broken at first order near the tops of the magnetic well, which unavoidably leads to a small, but finite, $r_a$ below which trapped particles near the magnetic well tops are not maximum-$J$. We discuss this issue further in Appendix~\ref{app:omnigeneity-breaking-wa}, but shall otherwise make the assumption of exactly omnigenity, so that this term may be neglected.
    \begin{figure}
        \centering
        \includegraphics[width=0.85\textwidth]{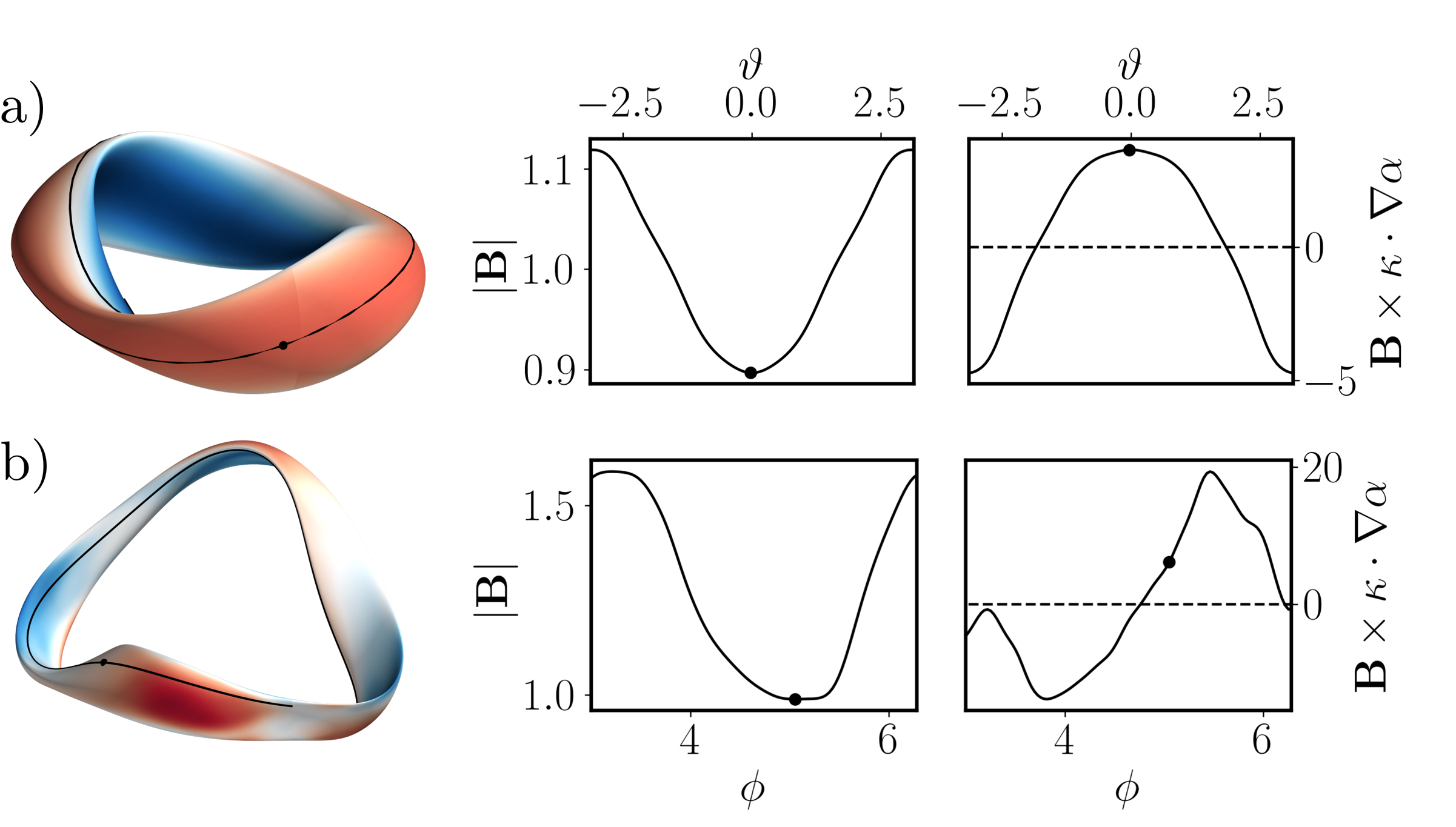}
        \caption{\textbf{Curvature about the minmum in a QS and QI field.} The plots show a comparison of the curvature ($\mathbf{B}\times\kappa\cdot\nabla\alpha$) in (a) a QA (the precise QA in \cite{landreman2022magnetic}) and (b) a QI (the $N=2$ configuration in \cite{goodman2023constructing}) configurations. The left plots show a 3D rendition of the boundary of the configurations, in which the colours display the quantity $\mathbf{B}\times\kappa\cdot\nabla\alpha$, with blue/red representing negative/positive values (good/bad curvatures) respectively. The plots to the right show the magnetic field $|\mathbf{B}|$ and the curvature along the field line, using the cylindrical coordinate $\phi$ and the PEST poloidal angle $\vartheta$ as coordinates along the fieldline. The scatter plot indicates the position of the minimum. These two examples illustrate the qualitative difference in the curvature parity between a QI and a QS field.}
        \label{fig:kappa-qi-qs}
    \end{figure}
    \item In QS stellarators, the fraction of trapped particles decreases as one approaches the magnetic axis, as it is the first-order poloidal variation of the field strength that defines the trapping well. In QI stellarators, this is not the case and there remains a finite trapped population on axis. This difference is important close to the axis, where there is a region with so-called \textit{potato-orbits} \citep{helander2005collisional,rodriguez2023precession} in QS stellarators where the thin-orbit approximation to $\mathcal{J}_\parallel$ breaks down.
    \item In QS, it is \textit{always} the case that there are trapped particles precessing in opposite directions. Deeply trapped particles precess like in a minimum-$J$ field ($q \omega_\alpha < 0$), while barely trapped ones do so in the opposite way. This is generally not the case in QI, where the function $f$, Eq.~(\ref{eqn:fun_f_nae_qi}), has more freedom than in the QS case. This stems from differences in the distribution of good and bad curvature about the minimum of $|\mathbf{B}|$. 
\end{enumerate}
\par
These conclusions hold rigorously on the magnetic axis and, by continuity, also in its vicinity. The differences in the behaviour of curvature can be seen to hold approximately true in practice, as is shown through two examples in Figure~\ref{fig:kappa-qi-qs}. In addition, in the QS case it has been shown that the near-axis description remains instructive as a model of QS configurations beyond its asymptotic regime (see Figure~\ref{fig:qs-wa}). 
\par
The most important observation is that, in the context of QI configurations, maximum-$J$ behaviour appears possible even in a vacuum field. Within the near-axis framework, whether this is the case depends on the sign of the bounce-average of the function $f(\varphi)$, which is analysed below. 

\subsection{Shallowly and deeply trapped particles in QI}
The special character of the bottom and top of the magnetic well provides us with a simple way to assess the possibility of maximum-$J$ behaviour in QI stellarators. 
\par
Satisfying the maximum-$J$-condition for the most shallowly trapped particles tends to be easy in practice. These particles spend most of their time close to the turning point where the field strength reaches its maximum along the field line, $B(\psi, \alpha, \varphi) = B_{\rm max}(\psi)$, which is independent of the field-line label $\alpha$ in a QI stellarator. Moreover, the maximum is located at a constant value of the toroidal Boozer angle, $\varphi = \varphi_{\rm max}$, which is independent of $\alpha$ \citep{Cary1997,landreman2012,helander2014theory}. At the maximum, the Taylor expansion of the field strength in the coordinates $\alpha$ and $\varphi$ is thus of the form
 \begin{equation}
    B(\psi,\alpha,\varphi) = B_{\rm max}(\psi) + \frac{B_{\varphi \varphi}}{2} (\varphi - \varphi_{\rm max})^2 + \cdots,  
    \label{eq:Bmax}
 \end{equation}
since, at the toroidal angle of the maximum, the following derivatives all vanish:
 \begin{equation}
    \frac{\partial B}{\partial \alpha} =  \frac{\partial B}{\partial \varphi} 
    =   \frac{\partial^2 B}{\partial \alpha \partial \varphi} =   \frac{\partial^2 B}{\partial \alpha^2} = 0.     
 \end{equation}
 Therefore $\nabla^2 B=\nabla^2 B_\mathrm{max}+B_{\varphi\varphi}|\nabla\varphi|^2+\dots$, where the second term is negative at the top of the well. Now, in a vacuum field (zero plasma pressure), the quantity $B^2=B_x^2 + B_y^2 + B_z^2$ is a subharmonic function since
    $$ \nabla^2 B^2 = \nabla \cdot \nabla (B_x^2 + B_y^2 + B_z^2) = 2( |\nabla B_x|^2 + |\nabla B_y|^2 + |\nabla B_z|^2) \ge 0, $$
where we have used $\nabla^2 B_x = 0$ etc. for each Cartesian component of $\bf B$ \citep{Solovev1970}. Because of the maximum principle for subharmonic functions, the maximum of $B^2$ over any closed, bounded domain cannot be attained in its interior unless $B^2$ is constant \citep[Thrm.~1, Ch.~6.4.1]{evans2022partial}. It thus follows that the function $B_{\rm max} (\psi)$ in Eq.~(\ref{eq:Bmax}) cannot have a local maximum at any value of $\psi$. Therefore, $B_{\rm max}'(\psi)\geq 0$ for all $\psi$ and, as a result of Eq.~(\ref{omegaa0}), shallowly trapped particles must satisfy the maximum-$J$-condition, $q\omega_\alpha > 0$, throughout the plasma.\footnote{Of course, $B'(\psi)$ could also vanish without $B(\psi)$ acquiring a local maximum, but we do not consider such singular cases that are not robust to arbitrarily small perturbutions.} It should be clear that this argument holds for any omnigenous magnetic configutation, including a quasisymmetric stellarator.
\par
The situation is very different at the opposite end of the trapped population. The most deeply trapped particles at the bottom of the well tend to precess in the unfavourable direction. Indeed, these particles frequently (though not always, see Figure~\ref{fig:qi-alan-examples-wa}), have the smallest value of $q\omega_\alpha$. The relevant question here is then whether this tendency be reversed to make the system acquire the maximum-$J$ property at the bottom of the trapping well. 
\par
In order to understand the maximum-$J$ condition on deeply trapped particles, we resort to Eq.~(\ref{eqn:deep-QI-wa-nae}), which provides an explicit expression for $\omega_\alpha$ in terms of $B_{20}$, $B_0$ and $d$. The expressions at this point may be further simplified by realising that the curvature of the magnetic axis vanishes at the point of minimum $|\mathbf{B}|$, i.e. $\kappa(\varphi) = 0$ when $\varphi = 0$. As a result, the behaviour of $|\mathbf{B}|$ is significantly constrained. The field around this point thus resembles a straight magnetic mirror, and thus we can gain some intuition about QI fields by analysing a straight magnetic mirror, which is done in Appendix~\ref{app:magMirror}.
\par
With these remarks in mind, we proceed to consider the governing set of equations in the near-axis framework. Formally, we explore the properties of a magnetic field in MHD equilibrium that satisfies the solenoidal condition, and possess flux surfaces at $\varphi=0$. The analysis is presented in Appendix~\ref{sec:appB20Deep}, assuming perfect QI at first order for simplicity, and leads to an explicit form for the average radial derivative of the magnetic pressure ($B_{20}$), which is primary ingredient of Eq.~(\ref{eqn:deep-QI-wa-nae}), explicitly constructed in Eqs.~(\ref{eqn:B20-min-exp}) in Appendix~\ref{sec:appB20Deep}. When the condition of omnigeneity at second order is also imposed (that is, the choice of functions is made so that QI is satisfied at the bottom of the well at second order), the function $f$ in Eq.~(\ref{eqn:wa-QI-approx}) can be written in terms in the following form 
% For an equilibrium field that is straight to leading order, it follows that
%     $$B_{20}=-\frac{\mu_0p_2}{B_0^2}+\sum_i \frac{\mathcal{P}_i}{4(l')^2},$$
% where
%     \begin{subequations}
%     \begin{gather}
%         \mathcal{P}_{B_0''}=-\frac{B_0''}{B_0}\frac{\bar{B}^2}{B_0^2\bar{d}^2}, \\
%         \mathcal{P}_{\Bar{d}\Bar{d}''}=\Bar{d}\Bar{d}''\left(1-\frac{\bar{B}^2}{B_0^2\bar{d}^4}\right), \\
%         \mathcal{P}_{\tau_0^2}=(\Bar{d}\tau_0l')^2\left(3-\frac{\bar{B}^2}{B_0^2\bar{d}^4}\right), \\
%         \mathcal{P}_{I_2}=-4(l')^2\bar{d}^2\frac{\tau_0I_2}{\bar{B}}.
%     \end{gather}
%     \end{subequations}
%     Here $\bar{d}=d/\kappa$.  
%     These can be interpreted (at least qualitatively) from the perspective of min-B mirrors, and the positive contribution of a local shear driven by an increase in elongation (and, of course, affected y torsion and current in the general case). The $B_0''$ term is simple to interpret \citep{furth1964closed,savenko2006mhd,aagren2004magnetic}. 
%     \par
    $$f_0=f_p+f_{B_0''}+f_{I_2}+f_{\tau_0^2}+f_\mathrm{QI},$$ 
with
\begin{subequations}
    \begin{gather}
        f_p=-\frac{\mu_0p_2}{B_0^2}, \\
        f_{B_0''}=-\frac{1}{2(l')^2}\frac{\sqrt{\bar{\alpha}}}{\bar{\alpha}+1}\frac{B_0''}{B_0}, \\
        f_{\tau_0^2} = \frac{\sqrt{\bar{\alpha}}}{1+\bar{\alpha}}\tau_0^2, \\
        f_{I_2}=-\frac{2\sqrt{\bar{\alpha}}}{1+\bar{\alpha}}\frac{I_2}{B_0}\tau_0, \\
        f_\mathrm{QI} = -\frac{1}{4B_0}\frac{2}{1+\bar{\alpha}}\left.\left(\frac{B_0^2d^2}{B_0'}\right)'\right|_{\varphi=0}, \label{eqn:f_QI}            
    \end{gather} \label{eqn:F-min-exp}
\end{subequations}
where $\bar{d}=d/\kappa$ and $\bar{\alpha}=\bar{d}^4B_0^2/\bar{B}^2$ is directly related to the elongation (along the direction of curvature) of the cross-section at the toroidal position $\varphi=0$. There are thus five different terms contributing to $f$, which we proceed to analyse individually. 
% But before doing so we note that $B_{20}$ itself has an additional dependence on $\bar{d}''$, and that only upon the consideration of omingeneity through second order does it disappear (although it must be finite to satisfy that omnigeneity condition).

\subsubsection{Role of pressure}
As expected, increasing the pressure gradient supported by the field, $|p_2|=(\bar{B}/2)|\mathrm{d}p/\mathrm{d}\psi|$, increases $f_0$. That is, it makes deeply trapped particles more likely to satisfy the maximum-$J$ condition $q\omega_\alpha>0$. This is the well-known effect of finite $\beta$ improving the maximum-$J$ property of QI fields, which is precisely the opposite to the na\"{i}ve interpretation of the role of the pressure from the explicit term in Eq.~(\ref{omegaa-simple}), which vanishes at the extremal points. This effect of $p'$ is the same at both the bottom and top of the well, thus bringing both deeply and barely trapped particles towards $q\omega_\alpha>0$. To a large extent, then, we would expect the rest of the trapped population to do likewise, and although we cannot prove it, it is convenient to think of this effect as an overall upshift of $q\omega_\alpha$. This diamagnetic behaviour turns out to be the same in the axisymmetric/quasisymmetric limit \citep[Eq.~(3.6a)]{rodriguez2023precession}\citep{Rosenbluth1971}.
\par
Whether a non-zero pressure gradient is needed to ensure $f_0>0$ depends on the relative size of the other terms in Eqs.~(\ref{eqn:F-min-exp}). At a minimum, we need to overcome the detrimental effect of being located at the minimum of the magnetic field, $f_{B_0''}$. The plasma beta, $\beta_0 = 2 \mu_0 p_0 / B_0^2$, that neutralises this term, $|f_p|=|f_{B_0''}|$, we define to be $\beta_J$ and provides an estimate of the critical $\beta$ that leads to maximum-$J$ behaviour. If we consider a simple quadratic pressure profile for a field with minor radius (defined as in the near-axis) $r=a$, then $\mathrm{d}p/\mathrm{d}\psi\approx -2p_0/B_0 a^2$, and thus $f_0>0$ when
\begin{equation}
    \beta_0>\beta_J\equiv R_M\frac{\sqrt{\bar{\alpha}}}{1+\bar{\alpha}}\left(\frac{a}{L_B}\right)^2, \label{eqn:beta_J}
\end{equation}
and $R_M/L_B^2=\partial_{\ell \ell} \ln B_0$, $R_M=(B_\mathrm{max}-B_\mathrm{min})/B_\mathrm{min}$ is the mirror ratio and $L_B$ the characteristic parallel length-scale of the bottom of the well. For typical field parameters (a moderately elongated cross-section with $\bar{\alpha}\sim 1$, a well of the scale of the major radius, and $R_M\sim0.2$), we get critical plasma betas around (or below) a percent. This suggests that maximum-$J$ behaviour is readily achievable at a moderate plasma $\beta$. 

\subsubsection{Difficulties at the minimum}
The special character of the bottom of the well is captured by the $f_{B_0''}$ term, which is negative (since $B_0''>0$) and thus opposes maximum-$J$ behaviour. The situation is however reversed at the maximum of $B$, where $B_0''<0$ and $f_{B_0''} > 0$, implying an `intrinsic' tendency for maximum-$J$ behaviour at this location. In fact, in a vacuum field, $f_0>0$, in agreement with the general proof presented in the previous sections. 
\par
The detrimental contribution of $f_{B_0''}$ at the bottom of the well can be mitigated by reducing $B_0''$ to a minimum, i.e. by flattening the bottom of the well (increasing $L_B$) or reducing the mirror ratio $R_M$. The elongation of the flux-surfaces may also be tweaked to reduce $f_{B_0''}$, which is maximal for circular cross-sections ($\bar{\alpha}=1$). However, note that the other geometric contributions to $f_0$, Eqs.~(\ref{eqn:F-min-exp}) also depend on the elongation, and thus in relative terms this shaping may not be as effective although it does reduce $\beta_J$. 
\begin{figure}
    \centering
    \includegraphics[width=0.8\textwidth]{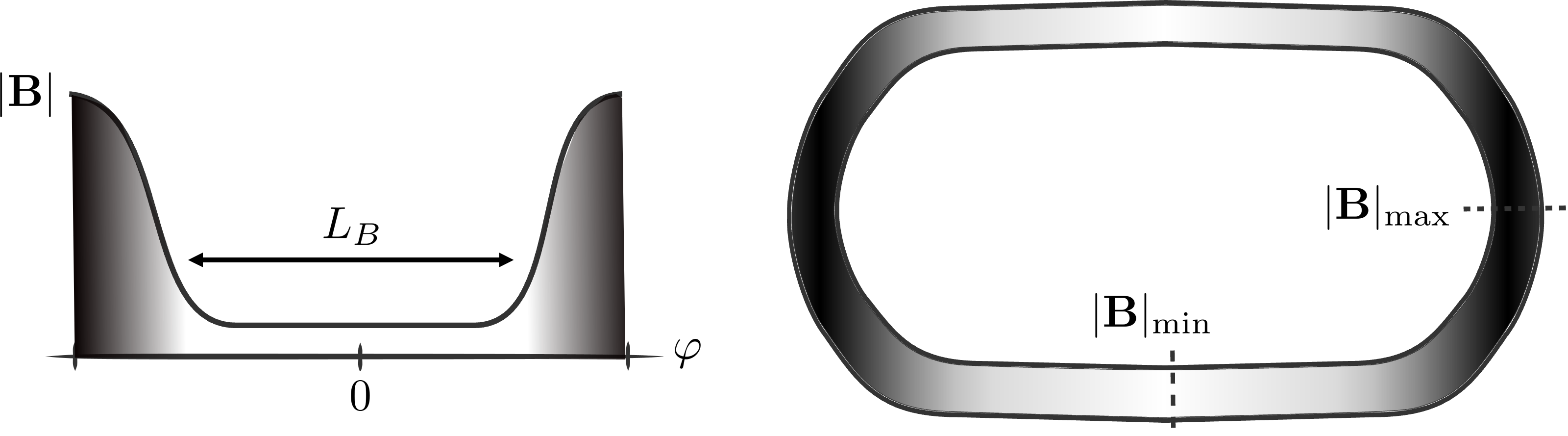}
    \caption{\textbf{Schematic depiction of trapping wells that favour maximum-$J$ behavior.} Certain magnetic field features make them more prone to maximum-$J$ behaviour than others. In particular, wide and flat magnetic trapping wells (left) at straighter sections (right) of the stellarator favour maximum-$J$. The diagram on the right is a schematic top-down view on an $N=2$ stellarator, with the shading denoting $|\mathbf{B}|$.}
    \label{fig:f-opt-B0-shape}
\end{figure}
\par
Flattening the bottom of the well comes with potential drawbacks in the form of sensitivity to deviations from omnigeneity. Having an extended region of small $\partial B_0 / \partial l$ makes the configuration more susceptible to error fields, as secondary shallow trapping wells may be created by small perturbations. Such wells can be seen in the near-axis description of $|\mathbf{B}|$, where the requirement of omnigeneity (and more particularly, that of poloidally closed $B$-contours) limits the behaviour of $B_1$ depending on how flat $B_0$ is. As analysed in detail in \cite{rodriguez2023higher}, for a field near the minimum described by $B_1\sim\varphi^v$ and $B_0'\sim\varphi^{u-1}$, if $B_0$ is too flat, $u\geq 2v$ (except $u=2,~v=1$), this will introduce defects in $|\mathbf{B}|$ that lead to losses of deeply trapped particles. In fields where this situation is avoided, that is, for $u<2v$, the quantity $f_\mathrm{QI}$ vanishes, Eq.~(\ref{eqn:f_QI}). Only in the special case of a first-order curvature zero ($v=1$) and a quadratic well ($u=2$) is this contribution finite. Because $d^2/B_0'$ vanishes at $\varphi=0$ and is positive for $\varphi>0$, its derivative must be greater than or equal to zero. Hence, $f_\mathrm{QI}\leq 0$ at the bottom of the well; that is, its contribution is detrimental. This makes the `standard' first-order curvature zero and quadratic-well field \citep{camacho-mata-2022} particularly unfavourable for maximum-$J$ behaviour. Conversely, making the section of the field where minimum $|\mathbf{B}|$ is located as straight as possible (larger $u$ and $v$) should be beneficial (see Fig.~\ref{fig:f-opt-B0-shape}).

\subsubsection{Role of local shear, twist and elongation}
The contribution of the torsion of the axis to $f_0$ is always beneficial since $f_{\tau_0^2}>0$. In fact, the larger the torsion, the larger $f_0$, and thus, the closer the behaviour of deeply trapped particles will be to the maximum-$J$ requirement. The role of torsion may be surprising, but is understandable from the perspective of a straight magnetic mirror. At the bottom of a straight magnetic mirror, attaining maximum-$J$ is only possible if magnetic field lines are locally twisted (see Appendix~\ref{app:magMirror} and Eqs.~(\ref{eqn:min-B-mirr-i})-(\ref{eqn:min-B-mirr-ii})), i.e., if they experience some form of left-right asymmetry, and in this way possess non-zero local magnetic shear. In the context of our QI stellarator, the appearance of the $\tau_0^2$ term is simply a statement of the necessity of this twist about the magnetic axis, which is the geometric meaning of torsion.
\par
\begin{figure}
    \centering
    \includegraphics[width=0.6\textwidth]{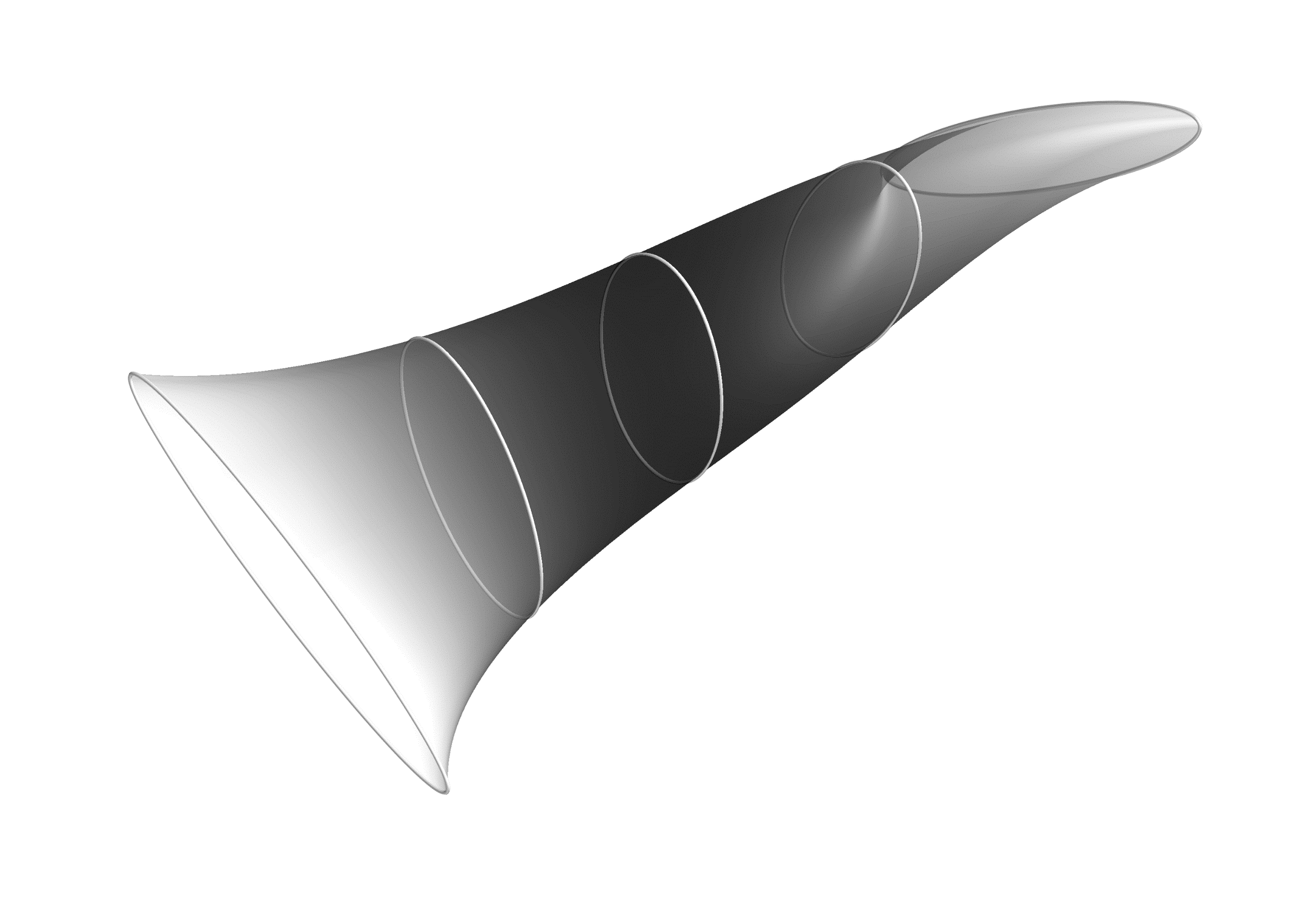}
    \caption{\textbf{Prototypical shape of flux surfaces of an omnigeneous, maximum-$J$ field near the point of minimum field strength.} 3D illustration of features generally expected from maximum-$J$, QI magnetic fields. Note that field lines twist and that the elongation grows with increasing distance from the field-strength minimum. }
    \label{fig:mirrorlike_behaviour}
\end{figure}

The comparison with a straight magnetic mirror is also helpful for understanding the role of flux-surface elongation, which in a mirror needs to increase away from the bottom of the magnetic well (see Figure~\ref{fig:opt-mirror-3d}). In the context of a stellarator-symmetric, QI stellarator, the elongation of the cross-section at the minimum is given by $\bar{d}^2$, as previously mentioned. Thus, a change in elongation would be expected to involve a term proportional to $\bar{d}''$.
\footnote{Identifying $\bar{d}''$ with the change in elongation at the minimum is not quite correct, as the elongation can vary through other means \citep{landreman2018a,rodriguez2023mhd} even when $\bar{d}$ does not change. In fact, one may show that at the minimum the change in elongation is affected by $\bar{d}''$, $B_0''$ and $\tau^2$. It can be rigorously shown that in the omnigeneous case increasing the torsion and $B_0''$ (when $\bar{\alpha}\neq1$ for the latter) always increase the elongation. Anyhow, the interpretation of $\bar{d}''$ as an addedd change to elongation is correct and illustrating.}
There is no term in $f_0$ which involves $\bar{d}''$, but this is a result of the condition of omnigeneity at second order. In fact, upon relaxing the latter, $B_{20}$ at the bottom of the well does depend on $\bar{d}''$, Eq.~(\ref{eqn:Ps-B20-min}), and it is only through the omnigeneity condition of Eq.~(\ref{eqn:omnnn-constr-min}) that this explicit dependence can be eliminated. From the omnigeneity condition it follows that $\bar{d}\bar{d}''(1+1/\bar{\alpha})=-(\tau_0\bar{d}l')^2(3+1/\bar{\alpha})+\dots$, meaning that increasing the torsion to favour $f_{\tau_0^2}$, requires one to modify $\Bar{d}''$ accordingly. The plasma cross-section must become elongated in the binormal direction away from the bottom of the well, which is a consequence purely of the omnigeneity condition. Looking at the contribution of the $\Bar{d}''$ term to the equilibrium equation of $B_{20}$, Eq.~(\ref{eqn:Ps-B20-min}), we see that $\bar{d}\bar{d}''(1-1/\bar{\alpha})>0$ promotes maximum-$J$ behaviour there. Shaping of elongation is once again vital, now for attaining maximum-$J$. For this shaping to be synergistic between the omnigeneity condition and maximum-$J$, we need a binormally elongated cross-section $\bar{\alpha}<1$ (i.e., binormal elongation at the minimum, $|\bar{d}|<1$). In practice, we seek shapes like the exagerated schematic in Figure~\ref{fig:mirrorlike_behaviour}, features that appear to be common in many QI configurations \citep{camacho-mata-2022, jorge2022c}.
\par
Given the central role played by the local field-line twist, a non-zero plasma current can be either beneficial or detrimental depending on its alignment with the torsion of the axis. As can be inferred from its contribution to the rotational transform \citep{MercierLuc1974}, a negative current acts constructively with a positive torsion (and vice-versa). The term $f_{I_2}$ reflects precisely this fact. In QI configurations, toroidal currents tend to be small and this contribution may then be disregarded. 
\par
All in all, we learn from this analysis that in a vacuum magnetic field there is only one way of possibly attaining maximum-$J$ behaviour for deeply-trapped particles, which is to have a large torsion at the point of minimum $B$, in the sense that $\tau_0 L_{B}>\sqrt{R_M/2}$, and thereby also a strong growth of binormal elongation. In other words, the binormal vector must rotate significantly within the magnetic well. From this analysis, it follows that it is possible to make deeply trapped particles precess in the favourable maximum-$J$ direction, unlike the case in quasi-symmetric or tokamak fields. This result is robust even if we acknowledge the impossibility of exact QI at first order in the near-axis expansion, because the violation of omnigeneity is only necessary near the maximum of $|\mathbf{B}|$. 
\par
The construction above poses some important practical difficulties. First of all, the field will generally tend to develop large shaping, both because the maximum-$J$ criterion requires significant torsion and omnigeneity also an increase in elongation (see the depiction in Figure~\ref{fig:mirrorlike_behaviour}). Comparing the various contributions to $f_0$ in Eq.~(\ref{eqn:F-min-exp}) reveals that the shaping needs to be specific and substantial to endow the deeply trapped particles with the maximum-$J$ property. If the shaping is constrained by practical considerations, as is usually the case in stellarator optimisation studies, the opportunity for attaining the maximum-$J$ property is correspondingly limited. How much shaping is sufficient varies from case to case, and the competition between the various field and geometric quantities (we shall recall that the shaping arguments above are a simplified local view near the minimum). The second thing to bear in mind is that our considerations here are limited to deeply (and barely) trapped particles. Intermediate orbits could behave differently, which is something that needs to be checked by computing $\omega_\alpha$ for all $k$.

\par

\subsubsection{Beyond stellarator-symmetry}
The realisation that a certain level of asymmetry about the bottom of the magnetic well provides the means to enhance the precession of deeply trapped particles opens the door to several possibilities. In the case of a straight mirror, one needs to break left-right symmetry in order to achieve an omnigeneous, maximum-$J$ field (see Appendix~\ref{app:magMirror} and \cite{Catto1981}). In the case of a QI stellarator it is then natural to ask the question whether breaking stellarator symmetry can be exploited to further improve the behaviour of the deeply trapped particles. 
\par
The procedure followed for the discussion above can be extended to the non-stellarator-symmetric case. The details of the derivation are presented in Appendix~\ref{sec:appB20Deep}. As a result of this extension, the expression for $f_0$ acquires a number of additional terms, and the stellarator-symmetric contributions are also modified. The additions are however limited if we specialise, for simplicity, to the case in which the elliptical cross-section at the bottom of the magnetic well remains up-down symmetric in the Frenet-Serret frame, i.e., if we choose $\sigma(\varphi=0)=0$. In that case, there is only a single additional term that arises from stellarator symmetry, which is proportional to $(\bar{d}')^2$,
\begin{equation}
        f_{(\bar{d}')^2} = \frac{(\bar{d}'/\ell')^2}{1+\bar{\alpha}}.
\end{equation}
It is clear that $f_{(\bar{d}')^2}\geq0$, and thus one can exploit stellarator-symmetry breaking to improve the maximum-$J$ behaviour of deeply trapped particles. In this particular case, $f_{(\bar{d}')^2}>0$ and thus any amount of symmetry breaking will help the precession of deeply trapped particles. When the level of asymmetry is such that it changes $\bar{d}$ significantly within the well, $L_{\bar{d}}<L_{B_0}$, where $L_{\bar{d}}^{-1}\sim \partial_\ell\ln \bar{d}$, this is capable of overturning the unfavourable contribution from $B_0''$ at the bottom of the well. This may be regarded as evidence that breaking stellarator symmetry might be beneficial.

\section{Some examples}
We showed in the previous section that it is possible to achieve maximum-$J$ behaviour for deeply trapped particles in QI stellarators but not in quasisymmetric ones. To attain this goal, the shaping of the magnetic field must however be carefully tailored and, in particular, significant shaping of the magnetic field is necessary to promote the correct precession behaviour. In this section, we construct concrete examples of QI fields and analyse their maximum-$J$ properties. 
\par
\subsection{Near-axis constructions}
We begin by considering configurations found through the near-axis expansion. To this end, we use the QI-specific developments of \cite{plunk2019direct} and \cite{rodriguez2023higher} as well as the general equilibrium framework of \cite{landreman2019}. To diagnose the resulting fields, we use the normalised precession frequency $\hat{\omega}_\alpha$, Eq.~(\ref{omegaa-hat}), where we express $\psi_a$, the value of the flux at the boundary, in terms of an effective aspect ratio. 
\par
To this end, we denote the length of the magnetic axis by
\begin{equation}
    2 \pi R = \oint dl,
\end{equation}
where the integral is taken along the axis once around the torus. To lowest order in the distance from the magnetic axis, the volume enclosed by a flux surface $\psi = \psi_a$ is equal to \citep{helander2014theory}
\begin{equation}
    V = 2 \pi \psi_a \oint \frac{dl}{B_0},
\end{equation}
where $B_0$ is the magnetic strength on axis. It is natural to define an ``average'' minor radius $a$ by setting $V = 2 \pi R \cdot \pi a^2$, and a logical definition of the aspect ratio is then
\begin{equation}
    A = \frac{R}{a} = \frac{1}{\pi} \sqrt{ \left( \oint dl \right)^3 \bigg\slash \left( 8 \psi_a \oint \frac{dl}{B} \right) },
\end{equation}
which may be expressed in terms of averages over the Boozer toroidal angle $\varphi$,
\begin{equation}
    \psi_a=\frac{1}{2}\left(\frac{G_0}{A}\right)^2\frac{\overline{1/B_0}^3}{\overline{1/B_0^2}},
\end{equation}
where the overline represents a toroidal average $\overline{(\dots)}=\int_0^{2\pi}(\dots) \mathrm{d}\varphi/2\pi$. We shall use this form of the toroidal flux in the presentation of numerical results to follow (choosing the representative value of $A\sim10$).  

\begin{figure}
    \centering
    \includegraphics[width=\textwidth]{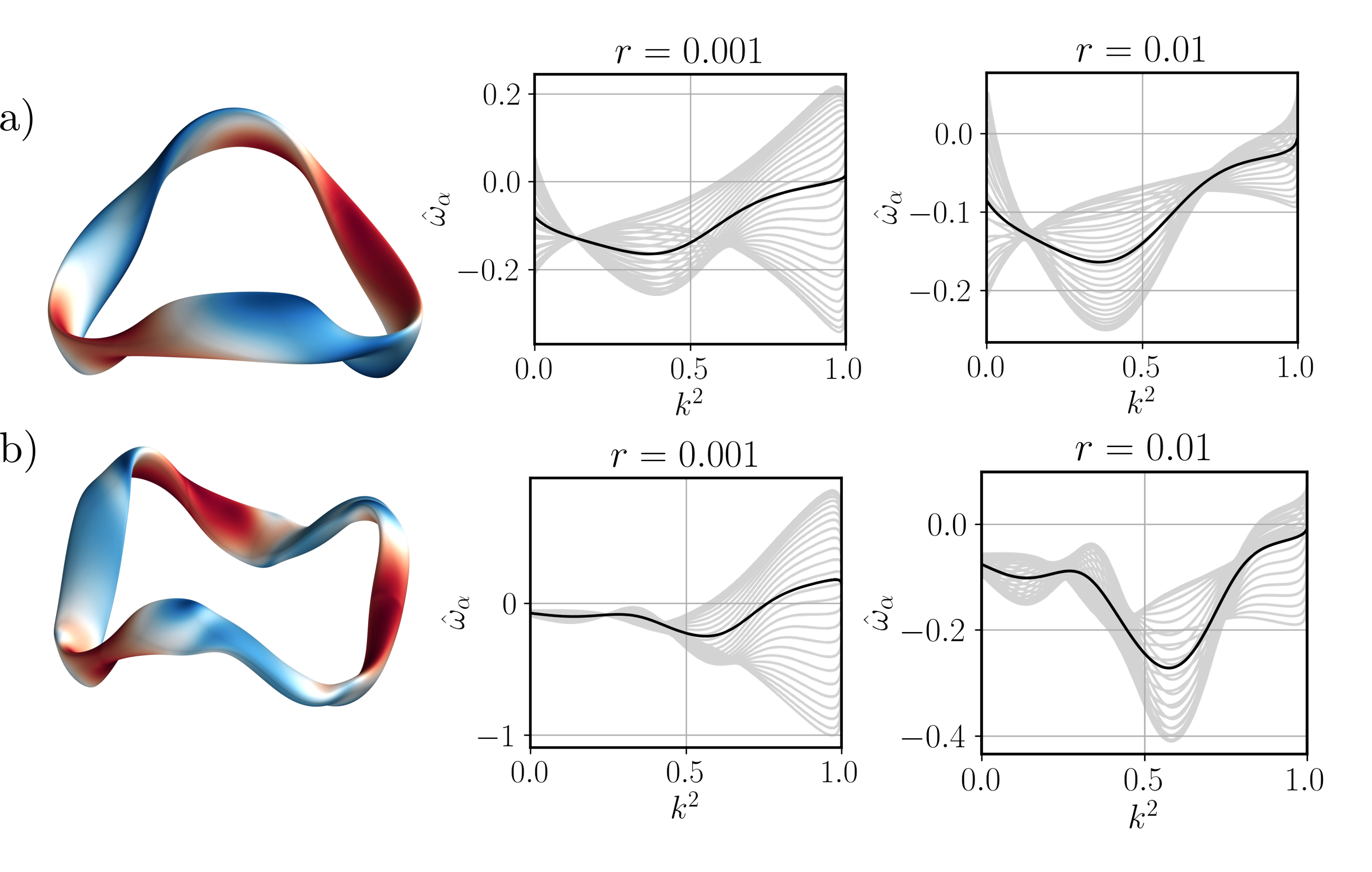}
    \caption{\textbf{Precession of trapped electrons for two near-axis QI examples.} 3D rendition of flux surfaces at $r=0.1$ (with the colour map representing the magnetic field strength) and normalised precession frequency $\hat{\omega}_\alpha$ at two different radii for near-axis fields with $N=2$ and $N=3$ of \cite{camacho-mata-2022}. The precession is computed using the analytical expressions derived in this paper. The near-axis fields have been constructed to second order, taking a `minimal shaping' construction $X_{2c}=0=X_{2s}$. The gray curves denote the variation of precession between different field lines (with the black curve corresponding to the average over $\alpha$), reflecting the non-omnigeneous nature of the fields. Two different origins of the variability are apparent: a roughly $r$ independent variation from the second order contribution, and a variation proportional to $1/r$ due to breaking omnigeneity at first order.}
    \label{fig:qi_analysis}
\end{figure}

To illustrate the behaviour of the precession in some near-axis constructions, we first consider configurations recently constructed by \cite{camacho-mata-2022}, which were designed to be quasi-isodynamic to first order in the distance from the magnetic axis. This work emphasised the reduction of shaping and neoclassical transport losses when a global equilibrium was constructed using the near-axis field. Because only first-order considerations were taken into account, there is in principle not a unique precession frequency characterising these configurations since a certain degree of freedom exists at second order to complete the construction.
Nevertheless, there is a `natural' second-order extension of these configurations, which we refer to as the \textit{minimal-shaping} construction, namely, the one that makes the $X_{2c}$ and $X_{2s}$ modulations in the near-axis vanish. The resulting field should be representative of the first-order construction, especially if one considers the construction of a global solution using the first-order fields. We construct the field following the equilibrium equations in \cite{landreman2019} (see Appendix~\ref{sec:appB20Deep}), using the code \texttt{pyQIC} \citep{pyQIC}. With such a second-order field in place, we may calculate the precession frequency, which is plotted in Fig.~\ref{fig:qi_analysis} for the $N=2$ and $N=3$ configurations of \cite{camacho-mata-2022}, where the necessary integrals were computed following Eq.~(\ref{eqn:qi-1/r-wa}) and (\ref{eqn:2nd-order-deviation-wa}). Some additional details are included in Table~\ref{tab:katia-qi-wa}.

\par
Similarly to the global QI-optimised stellarators in Fig.~\ref{fig:qi-alan-examples-wa}, these configurations do not have the maximum-$J$ property. Most of the trapped particles precess in the diamagnetic direction, but some particles with turning points close to the magnetic-field maxima behave as expected in a maximum-$J$ field. This behaviour is similar to that in axisymmetric/quasisymmetric fields: deeply trapped particles tend to co-precess with the diamagnetic frequency, while barely trapped ones do the opposite. The configurations importantly exhibit a significant field-line-to-field-line variation (see the different grey lines), as a result of deviations from omnigeneity. This should not come as a surprise given that the second-order construction is not QI. But in addition, as noted in the construction of \cite{camacho-mata-2022}, the tops of the wells deviate from QI already at first order. The result is a variability of $\hat{\omega}_\alpha$ that diverges as $r \rightarrow 0$ and primarily affects shallowly trapped particles, see Fig.~\ref{fig:qi_analysis}. Note, however, that this first-order effect is only noticeable very close to the axis. This is testimony to the quality of the QI optimisation performed by \cite{camacho-mata-2022}.
\par
Although the precise form of the curves in Fig.~\ref{fig:qi_analysis} depends on how the near-axis magnetic field has been completed at second order, the behaviour of deeply trapped particles is independent of this detail, as we learnt in previous sections. For these particles, it suffices to compute the various terms that make up $\omega_\alpha$ in Eq.~(\ref{eqn:deep-QI-wa-nae}) at the minimum of $B_0$. For this calculation, though, one should not use the form of $f_0$ in Eq.~(\ref{eqn:F-min-exp}) but instead relax the assumption of QI at second order. Setting the explicit non-QI contributions aside for now, we must use the expression for $B_{20}$ needed for Eq.~(\ref{eqn:deep-QI-wa-nae}). The necessary expressions are given in Appendix~\ref{sec:appB20Deep}, Eqs.~(\ref{eqn:Ps-B20-min}), where we use the notation $\mathcal{P}_i$ to denote the contribution of a quantity $i$ to $B_{20}$, in analogy to $f_i$ in $f_0$. These terms can be used to assess the near-axis construction at first order as in Table~\ref{tab:katia-qi-wa}. As $\bar{d}''=0$ (and thus $\mathcal{P}_{\bar{d}\bar{d}''}=0$) and $\bar{d}<1$ in these examples, it is a priori clear that it is \textit{impossible} for the deeply-trapped particles to have the maximum-$J$ property in these configurations. A finite plasma $\beta$ would be necessary to attain this property. Much like the critical $\beta_J$ in Eq.~(\ref{eqn:beta_J}), we can define $\beta^\star=a^2(\sum\mathcal{P}_i)/(2(\ell')^2)$, where $a$ is the value of $r$ at the plasma edge, which we take to be at roughly $a\sim R/10$. The quantity $\beta^\star$ represents the plasma $\beta$ necessary to make the vacuum QI configuration reverse the behaviour of deeply trapped particles. The key features for the equilibria in Figure~\ref{fig:qi_analysis} are collected in Table~\ref{tab:katia-qi-wa}. We have at this point neglected the contribution of $\omega_\alpha^\mathrm{non-QI}$ to Eq.~(\ref{eqn:deep-QI-wa-nae}), and since the configurations listed in Table~\ref{tab:katia-qi-wa} are not exactly omnigenous, it is actually necessary to increase $\beta$ further in order to overcome the $\alpha$-dependence. The requisite $\beta$ can be estimated from Eqs.~(\ref{eqn:wa-non-qi-deep}) and (\ref{eqn:B2c-deep}), and we may thus define $\beta^\star_\mathrm{tot}$ as the total $\beta$ needed to make the `least-maximum-$J$' field line be so.
\begin{table}
        \centering
        \begin{tabular}{c|c|c|c|c|c|c|c}
             & $\bar{\alpha}$ & $\mathcal{P}_{B_0''}$& $\mathcal{P}_{\tau_0^2}$& $\mathcal{P}_{\Bar{d}\Bar{d}''}$& $\mathcal{P}_\mathrm{QI}$& $\beta^\star$ & $\beta^\star_\mathrm{tot}$ \\\hline
            % QI nfp1 r2 & 0.040 & -0.13 & -1.00 & 0 & -0.20 & 1.9\% \\
            % QI nfp2 r2 & 0.12 & -0.34 & -1.53 & 0.45 & -1.16 & 4.0\% \\
            N2 & 0.21 & -1.83 & -10.48 & 0 & -8.82 & 5.5\% & 15.5\%\\
            N3 & 0.16 & -10.01 & 0 & 0 & -17.14 & 4.3\% & 4.7\% \\
        \end{tabular}
        \caption{\textbf{Critical $\beta$ and geometric contributions.} Geometric parameters and critical $\beta$ for the $N=2$ and $N=3$ near-axis QI examples in \cite{camacho-mata-2022}. Here $\beta^\star$ is a measure of the required plasma beta to prevent deeply trapped particles from precessing in the diamagnetic direction in an idealised omnigeneous field ($\beta_\mathrm{tot}^\star$ if the non-omnigeneous nature of the field is considered).}
        \label{tab:katia-qi-wa}
    \end{table}

\par
\begin{figure}
    \centering
    \includegraphics[width=0.7\textwidth]{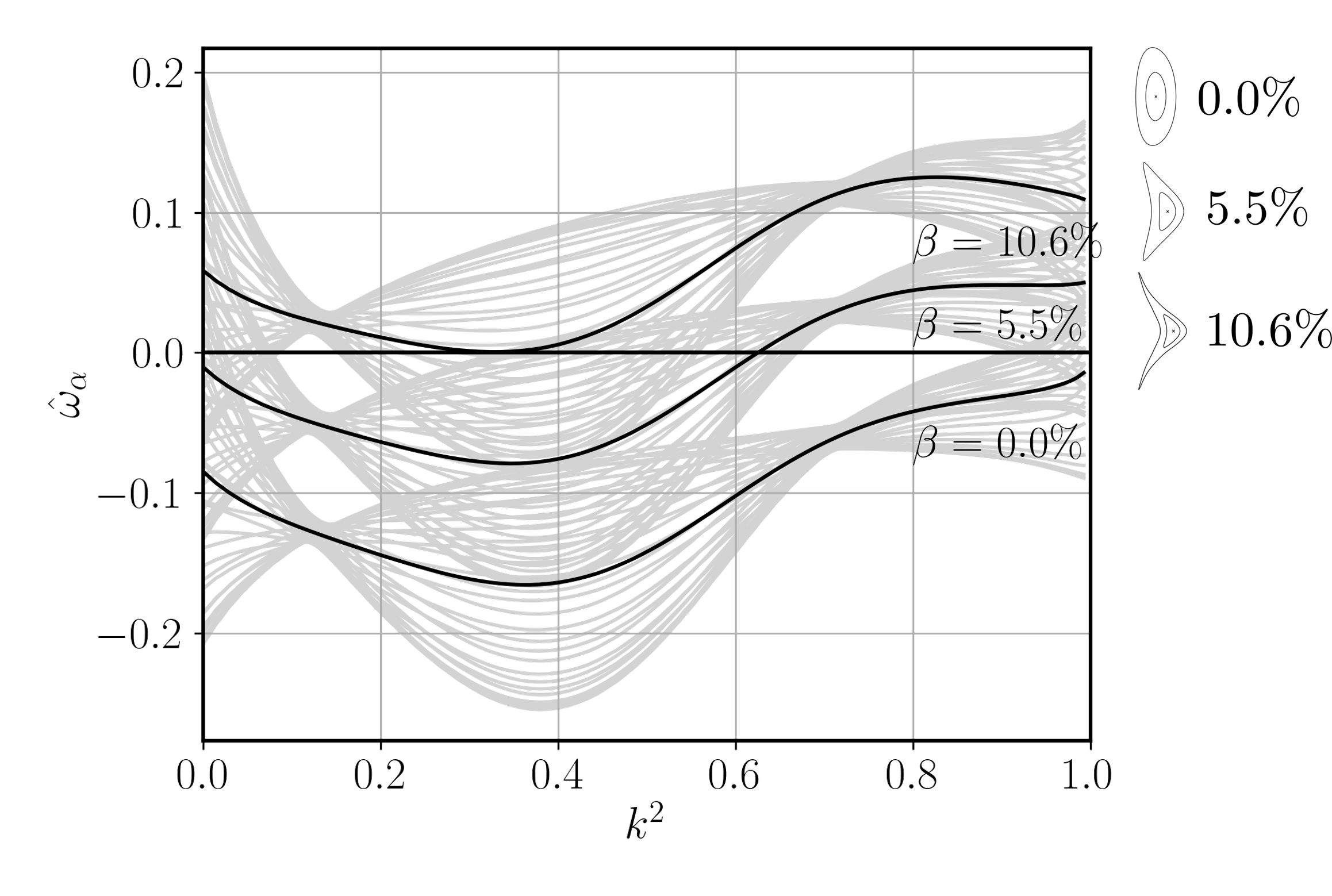}
    \caption{\textbf{Reversal of trapped-particle precession with increasing $\beta$.} Normalised precession frequency $\hat{\omega}_\alpha$ for three different values of plasma $\beta$ for the `minimally-shaped' near-axis field $N=2$ from \cite{camacho-mata-2022} at $r=0.01$. The black lines correspond to the $\alpha=\pi/4$ field line, and gray curves reflect the variation in precession frequency due to non-omnigeneity. The analytical estimate for the normalised pressure at which deeply trapped particles reverse their precession is $\beta^\star = 5.5\%$ and corresponds to the middle set of curves. The legend on the right shows how the plasma cross section at the radii $r=0.05,~0.1$ and toroidal angle $\phi=0$ changes with $\beta$.}
    \label{fig:qi_beta}
\end{figure}

The analytical value of $\beta^\star$ can be seen to be correct in the example of Figure~\ref{fig:qi_beta}, where the plasma $\beta$ of the configuration in Fig.~\ref{fig:qi_analysis}a increases. Here, the pressure gradient has been varied whilst the shape of the magnetic axis and the ellipticity of the flux surfaces in its vicinity is kept fixed to first order. Note that different authors and contexts mean different things by `increasing $\beta$' (i.e., different features of the equilibrium are kept constant), making direct comparison difficult. Of course, reaching this $\beta$ is necessary but not sufficient for maximum-$J$ of the whole trapped population. This is especially true in the examples in Fig.~\ref{fig:qi_analysis}, where it is not the most deeply trapped particles that have the largest precession frequency. From the preceeding analysis, we expect a non-zero plasma $\beta$ to introduce an overall upshift of the precession $q\omega_\alpha$, which is indeed seen numerically. 

\par
    % \begin{figure}
    %     \centering
    %     \includegraphics[width=\textwidth]{figures/wa_nae_example.png}
    %     \caption{\textbf{Precession of trapped electrons for two near-axis QI examples.} The plot shows the 3D rendition of the flux surface at $r=0.1$ (with the colormap representing the magnetic field magnitude) and the precession $\omega_\alpha/H$ for two radial positions for the near-axis fields (a) QI nfp1 r2 and (b) QI nfp2 r2 in Table~\ref{tab:my_label}. The gray curves denote the variation of precession between different field lines, a measure of the non-omnigeneous nature of the fields. This shows the two possible origins of the variability: a roughly $r$ independent variation from the second order contribution, and a $1/r$ growing variation due to breaking omnigeneity at first order. The increase in that occurs in the latter is clear in (b).}
    %     \label{fig:qi_analysis}
    % \end{figure}

\par
A central conclusion of our analysis thus far is the possibility of making deeply trapped particles acquire maximum-$J$ behaviour without the need of a nonzero plasma pressure. Accordingly, we now attempt to construct such a field first through optimisation within the near-axis framework. We construct an optimisation measure for maximum-$J$ by summing $\omega_\alpha^2$ (which we have learnt how to compute) over values of $k$ that satisfy $q \omega_\alpha<0$, call it $g_{\omega_\alpha}$. We are also interested in imposing the condition of QI, especially at second order. At the maxima and minima of the field strength we learnt in this work that we must satisfy Eq.~(\ref{eqn:omnnn-constr-min}), from which we may construct an additional cost function, call it $\check{g}_\mathrm{QI}$. Under the assumption of satisfying this condition, one can show that it is possible to choose the near-axis construction at second order in such a way that it guarantees the correct second order QI behaviour elsewhere. We give the most essential elements of this in Appendix~\ref{sec:app_opt_nae}, but leave a full exploration to a later publication. Note that this way of completing the solution is formally correct, but in practice (i.e., when taking into account shaping, QI breaking in buffer regions, etcetera) it may not be the best choice. For our proof-of-principle, though, it should suffice. With this, then, we construct our near-axis cost function simply as the weighted sum of the negative $q\omega_\alpha$ and the QI condition, Eq.~(\ref{eqn:omnnn-constr-min}), at the $|\mathbf{B}|$ extrema, $g=g_{\omega_\alpha}+\check{g}_\mathrm{QI}$. 
\par
\begin{figure}
            \centering
            \includegraphics[width=0.8\textwidth]{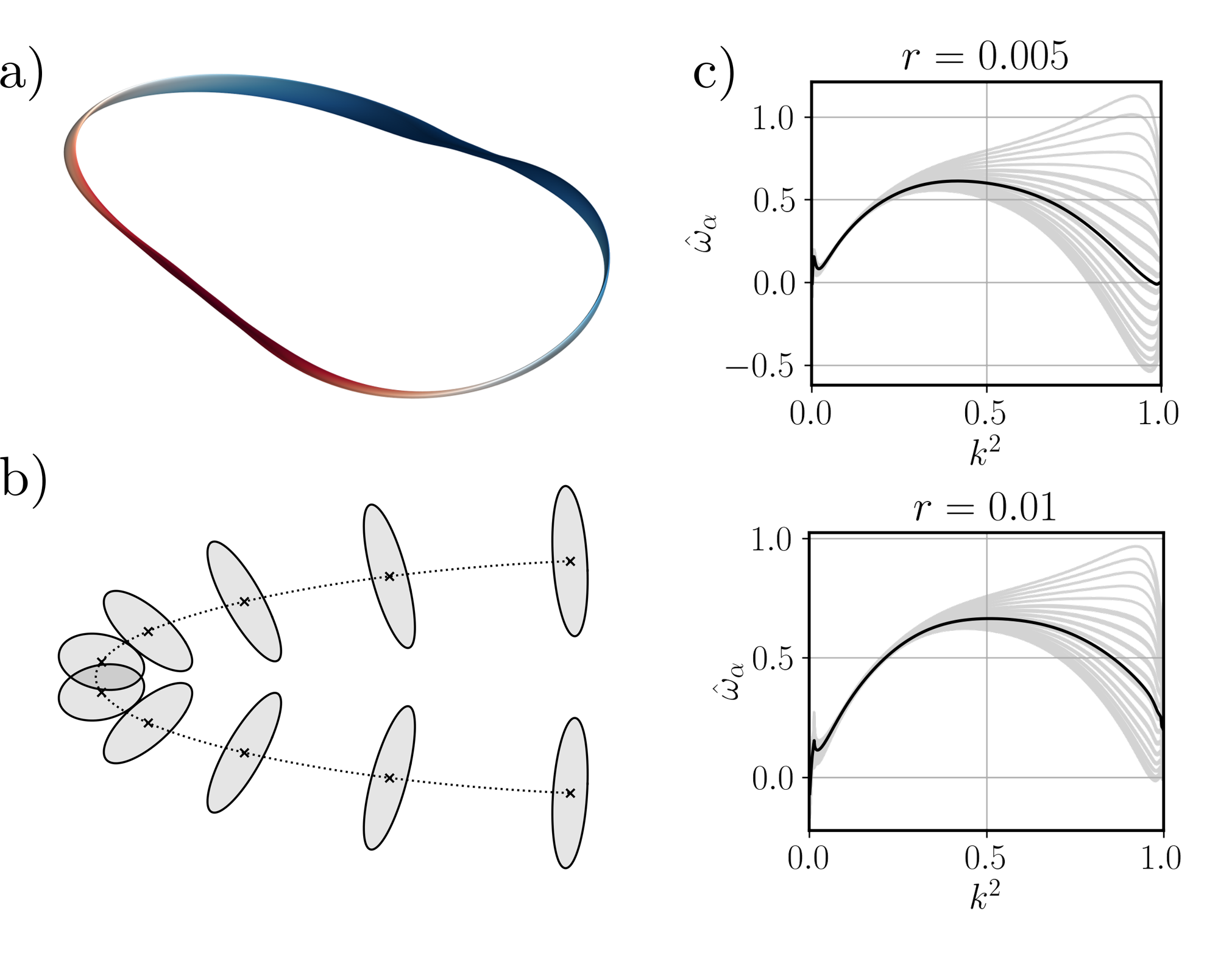}
            \caption{\textbf{Proof-of-principle of a maximum-$J$-optimised near-axis field.} The plots show a near-axis field optimised so as to exhibit quasi-isodynamicity and maximum-$J$ behaviour, especially near the minimum of the magneic field on each flux surface. The shaping of the configuration is very large, and thus the resulting field impracticable, but shows that the optimisation criteria can be met. a) A 3D rendition of the field for $r=0.02$, using only the 1st order description. (Second-order contributions are large and obscure the visualisation.) b) First-order cross sections near the minimum of $|\mathbf{B}|$ (in $(R,Z)$ coordinates and between $\phi=3\pi/4,~5\pi/4$). The dotted  curve represents the position of the magnetic axis. c) Precession $\hat{\omega}_\alpha$ computed numerically using $|\mathbf{B}|$ from the near-axis-expansion at two different radii on a number of field lines (gray curves, with the black representing the average). The increase in variability at low $r$ is due to the buffer region in which omnigeneity is broken. For this example, this contribution was not minimised. Close to the magnetic axis, the field satisfies the maximum-$J$ criterion for almost every orbit, except those trapped in secondary minima \citep{rodriguez2023higher}.}
            \label{fig:opt_nae_maxj}
        \end{figure}

With the cost function thus defined, we must explicitly state which our minimal degrees-of-freedom are. We shall allow only the axis shape and the function $\Bar{d}(\varphi)$ to vary, while keeping the field strength $B_0(\varphi)$ fixed. The idea is not to find a practical field, which would require limiting the shaping and other additional practical features (and unlocking other degrees of freedom such as $B_0$ and the order of curvature zeroes). We are simply aiming at the construction of a proof-of-principle field that exhibits maximum-$J$ behaviour in vacuum, especially for the deeply trapped particles. Other details of the optimisation are left to Appendix~\ref{sec:app_opt_nae}. In Figure~\ref{fig:opt_nae_maxj} we present the resulting optimised configuration for $N=1$. The shaping is forbiddingly large, owing to the fact that no attention was paid to limit it. Flux surfaces are extremely shaped and limit the physical radius of the configuration. However, we may from this approach formally construct an asymptotic $B$ at any $r$ (as a model if one wishes, $r_c\approx0.002$ \citep{landreman2021a}), and represent the geometry to first order to exhibit the features resulting from the optimisation. 
\par
The optimised field is one in which the deeply trapped particles exhibit maximum-$J$ behaviour (see Fig.~\ref{fig:opt_nae_maxj}c).\footnote{There exists some noise very close to the minimum owing to the appearence of `puddles' in $|\mathbf{B}|$ \citep{rodriguez2023higher}. This is the result of the simple axis shapes considered for this case. The noise should however disappear as $r\rightarrow0$.} To achieve this, the optimiser has found a field with the features identified in the preceding section. Specifically, the flux surfaces are twisted in the necessary manner and elongated away from the region of low magnetic field strength. Although maximum-$J$ behaviour has thus been achieved for deeply trapped particles, it is clear that the field is is not exactly omnigenous especially at larger $k$. This is mainly due to the first order deviations from omnigeneity which the optimisation target did not include (clearly seen by the growth of the variation going from the bottom to the top plot of Fig.~\ref{fig:opt_nae_maxj}). Reducing the `buffer region' (the region where omnigeneity is violated at first order) would reduce number of trapped-particles that do not satisfy the maximum-$J$ condition. 
        % \begin{figure}
        %     \centering
        %     \includegraphics[width=\textwidth]{figures/optMaxJconfig.png}
        %     \caption{\textbf{Max-J optimised 2nd order QI configuration.} The plots show a rendering of the flux surface (first order construction), cross sections, and the precession plots like before. The configuration was exclusively optimised for QI at second order and max-J using the full $\hat{k}$ population (although starting from a case optimised first using the simpler construction). It looks basically like a racetrack. $\iota_0=-0.41$.}
        %     \label{fig:my_label}
        % \end{figure}

\subsection{Traditionally optimised equilibrium}
Motivated by the theoretical possibility of a vacuum, QI, maximum-$J$ field, and proceeding beyond the near-axis expansion, we now turn to more traditional, global, stellarator optimisation. 
To this aim, we employ three target functions: $g_{QI}$ ensures that the field is QI, $g_{B_\mathrm{min}}$ ensures that the most deeply-trapped particles satisfy the maximum-$J$ criterion, and $g_{J}$ ensures that the other particles also do so.
The total target function that we minimise is thus
\begin{equation}
    g = g_{QI} + g_{B_\textrm{min}} + g_{J}. \label{eqn:opt_metric}
\end{equation}

Both $g_{QI}$ and $g_{J}$ are complicated targets, which are explained in greater detail by \cite{goodman2024}, and so is the starting point for the optimisation. Broadly speaking, $g_{QI}$ penalises the difference between $J$-contours on a flux surface by computing the difference in the second adiabatic invariant and a closely-related, artificially-constructed, perfectly QI flux surface. From the calculation of $\mathcal{J}_\|$, we also evaluate the term $g_{J}$, which imposes $\partial_\psi J < 0$ in this constructed field. Thus, as $g_{QI}$ and $g_J$ decrease, the field becomes more QI \textit{and} more maximum-$J$.

The term $g_{B_\textrm{min}}$ is a more straightforward target function. Using the fact that the maximum-$J$ condition corresponds to minimum-$B$ (for the most deeply- and shallowly-trapped particles), we simply designed this function to encourage the flux-surface's minimum field strength, $B_\textrm{min}(s)$, to have a positive derivative, where $s = (r/a)^2$. To do this, for every consecutive pair of flux surface, the target calculates $B_\textrm{min}$ on $s_0$ and $s_1=s_0 + \delta s$, and then the fractional difference between the two 
\begin{equation}
    \delta_B(s_0) = \frac{1}{\delta s}\frac{B_\textrm{min}(s_1) - B_\textrm{min}(s_0)}{B_\textrm{min}(s_1) + B_\textrm{min}(s_0)}.
\end{equation}
We can thus define the target as $f_{B_\textrm{min}} = \max(0.01 - \delta_B, 0)^2$, where the value of 0.01 has been chosen as an arbitrary positive number here.

\begin{figure}
    \centering
    \includegraphics[width=0.8\textwidth]{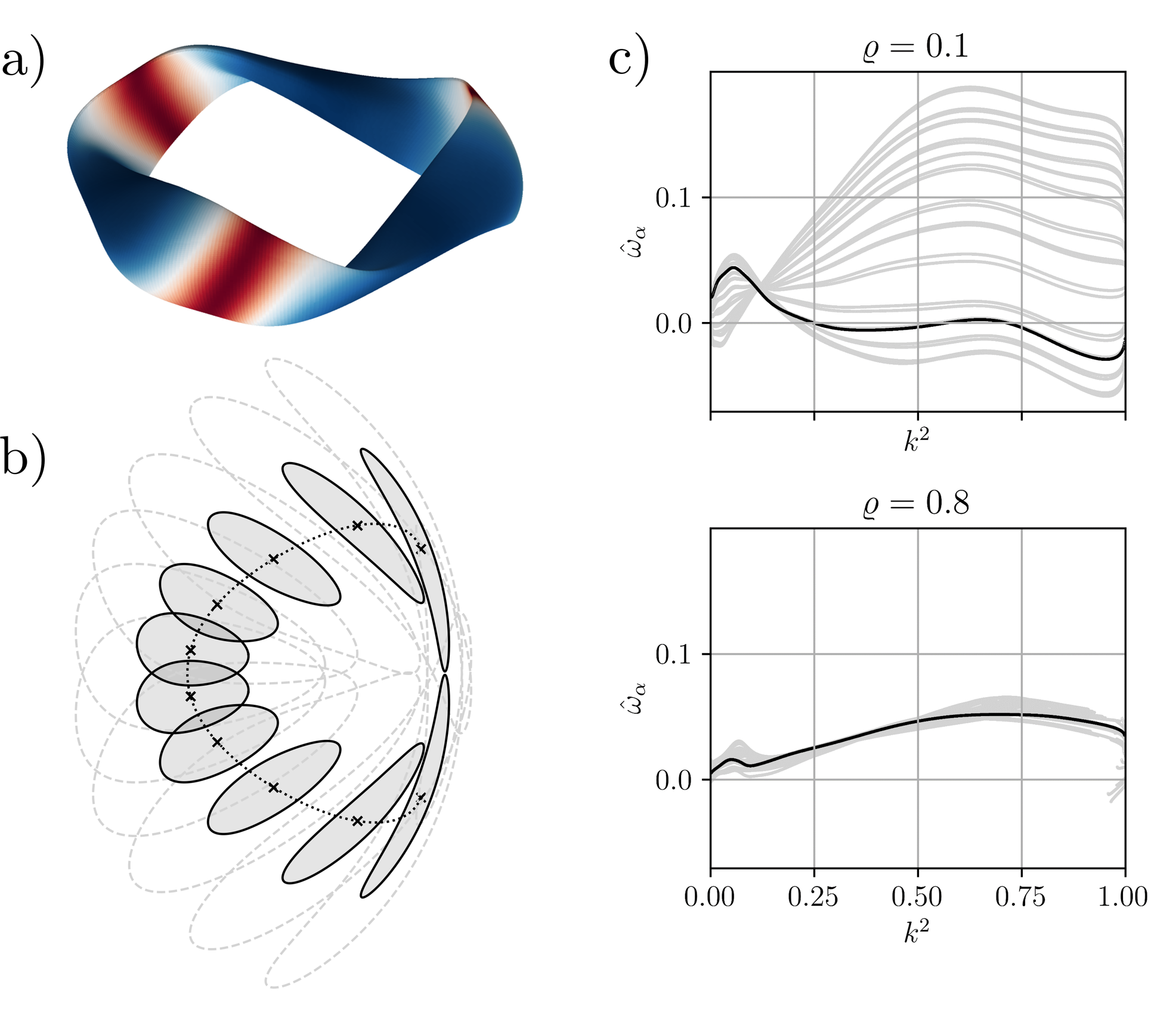}
    \caption{\textbf{Example of a maximum-$J$-optimised global equilibrium.} The plots show a magnetic field optimised by minimising Eq.~(\ref{eqn:opt_metric}) so as to attain quasi-isdynamicity and maximum-$J$ behaviour in a vacuum magnetic field. The configuration has an aspect ratio of $A_\mathrm{VMEC}\sim7$, three field periods, and exhibits strong shaping, which was not constrained in this proof-of-principle example. a) 3D rendition of the outermost surface of the field, where the colour map represents $|\mathbf{B}|$. b) Detail of cross-sections near the core (at $\varrho=0.1$, and for reference to indicate the large shaping of surfaces $\varrho=0.5$ as broken contours) and about the minimum of $|\mathbf{B}|$ (in $(R,Z)$ coordinates and between $\phi=3\pi/4,~5\pi/4$), showing features of twist and shape studied analytically. The dotted curve represents the position of the magnetic axis. c) Precession frequency at two radii on a number of field lines (gray curves, with the black $\alpha=\pi/2$). The increase in variability at low $r$ is a consequence of the breaking of omnigeneity. The overwhelming majority of all trapped particles satisfy the maximum-$J$ criterion, $\hat \omega_\alpha > 0$.}
    \label{fig:maxJ_vmec}
\end{figure}

\begin{figure}
    \centering
    \includegraphics[width=\textwidth]{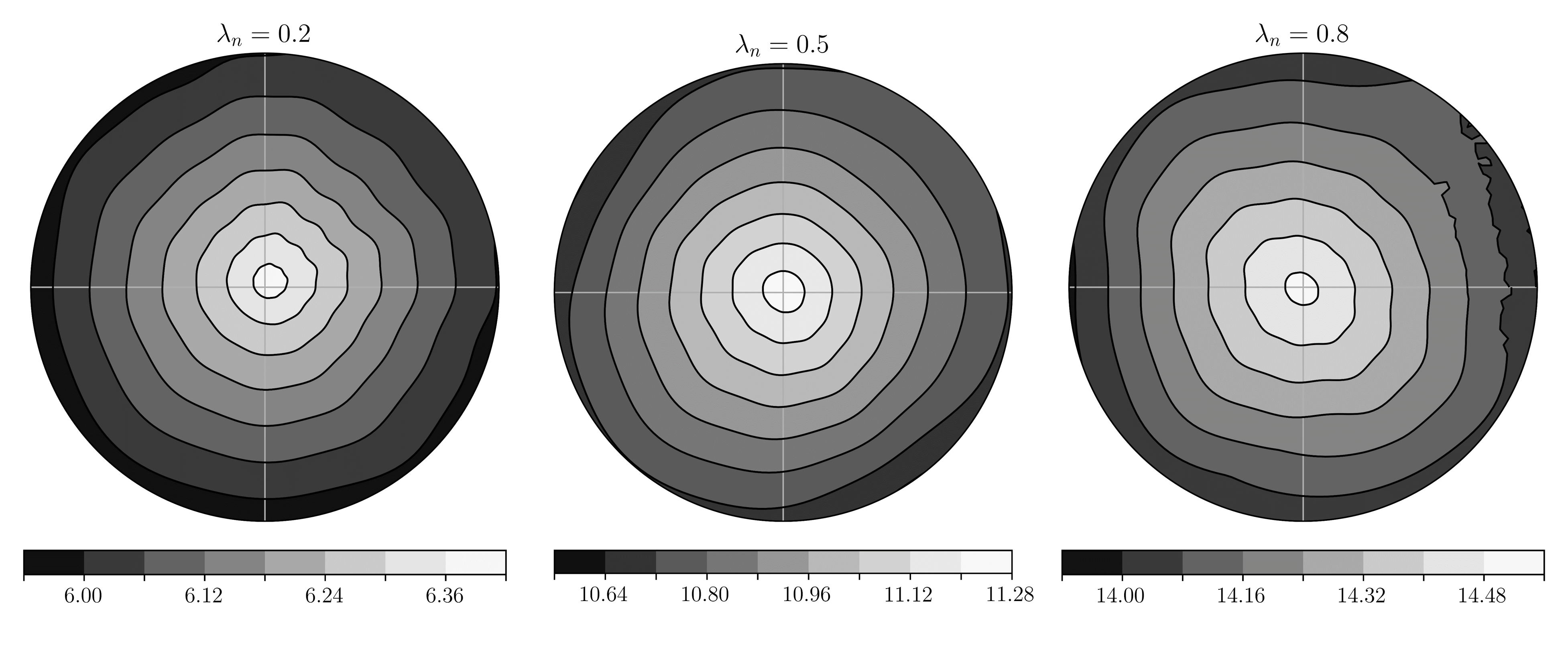}
    \caption{\textbf{Contours of the second adiabatic invariant $\mathcal{J}_\parallel$ in polar coordinates $(s,\alpha)$.} Contours showing the second adiabatic invariant as a function of the polar coordinates $(s,\alpha)$ for three different trapped particle classes. The values of $\lambda$ used are $\lambda=[(B_\mathrm{max}-B_\mathrm{min})\lambda_n+B_{min}]^{-1}$ where $B_\mathrm{max}$ and $B_\mathrm{min}$ denote the maximum and minimum field strength on the flux surface in question. An ideal omnigeneous field would have concentric circular contours. The wiggles in the contours is indicative of QI-breaking, which is particularly prominent close to the trapped-passing boundary (small $\lambda$). An ideally maximum-$J$ field would show a monotonic decrease of $\mathcal{J}_\parallel$ along any ray emanating from the origin. }
    \label{fig:J_cont_maxJ_vmec}
\end{figure}

We initialise optimisation from a near-axis construction with a flatter bottom of the magnetic well (which we know from the work above that should favour maximum-$J$), using techniques to be presented by \cite{plunk2024}. The resulting optimised configuration is indeed (mostly) maximum-$J$ and QI. This can be seen in the second adiabatic invariant contours of Figure~\ref{fig:J_cont_maxJ_vmec}, which show that $\mathcal{J}$ decreases from the centre of the $(s,\alpha)$ plot radially outward, and the contours are approximately circular. It is evident though that for trapped particles sufficiently close to the bottom of the wells (see left-most plot), there is a fraction of the population that does not behave in a maximum-$J$ fashion. This is signalled here by a relative shift of the approximately circular contours respect to the centre. As a result, there is a portion of field lines that are minimum-$J$ (see also the top plot of Figure~\ref{fig:maxJ_vmec}c). In the spirit of a more quantitative measure, it would be convenient to come up with a single scalar that indicates what `fraction' of the configuration truly behaves in a maximum-$J$ fashion. For a Maxwellian distribution function, the fraction of the trapped-particle population whose rotation satisfies $q\omega_\alpha>0$ is equal to 
\begin{equation}
    f_\mathrm{max-J}=\frac{1}{\mathcal{N}}\int_0^1\varrho \mathrm{d}\varrho \int_0^{2\pi}\mathrm{d}\alpha\int_{1/B_\mathrm{max}}^{1/B_\mathrm{min}}\Theta[q\omega_\alpha(\lambda, \alpha, \varrho)]\hat{\tau}_b(\lambda, \alpha, \varrho)\mathrm{d}\lambda,
\end{equation}
where $\Theta$ is the Heaviside step function, $\varrho=\sqrt{\psi/\psi_a}$,
\begin{equation}
    \mathcal{N} = \int_0^1\varrho \mathrm{d}\varrho \int_0^{2\pi}\mathrm{d}\alpha\int_{1/B_\mathrm{max}}^{1/B_\mathrm{min}}\hat{\tau}_b(\lambda, \alpha, \varrho)\mathrm{d}\lambda,
\end{equation}
and
\begin{equation}
    \hat{\tau}_b=\int_{\ell_L}^{\ell_R}\frac{\mathrm{d}\ell}{\sqrt{1-\lambda B}}
\end{equation}
is proportional to the bounce time of trapped particles. This way each flux surface is considered equally important (the Maxwellian distribution does not change from surface to surface). With this definition, for the optimised configuration above, $f_\mathrm{max-J}\approx0.996$. That is, a $99.6\%$ of the volume is maximum-$J$; i.e., for all intents and purposes, the configuration is maximum-$J$. The omnigeneous nature of the configuration can be quantified by the effective ripple \citep{nemov1999evaluation}, which measures the $1/\nu$ neoclassical transport and in the present case remains in the range $\epsilon_\mathrm{eff}\sim 0.15\%-0.45\%$ across the volume.

The resulting configuration also has a significant so-called ``vacuum magnetic well'' of about 3.3\% (as defined in \cite{landreman2022magnetic}), a property that is important for MHD stability \citep{landreman2021a}. A field is said to have a vacuum magnetic well if $V''(\psi) < 0$ (where $V$ is the volume enclosed by a flux surface). Generally, unless special care is taken, optimised configurations tend to have a ``magnetic hill'', $i.e.$ $V''(\psi)>0$, so it is notable that this configuration has such a substantial vacuum well.

The condition for a magnetic well means that, on a suitable average, the magnetic field strength should increase with minor radius. This can be seen from the circumstance that the volume enclosed by a flux surface $\psi$ is 
\begin{equation}
    V(\psi) = \int_0^\psi d\psi' \int_0^{2 \pi} d\theta \int_0^{2 \pi}  \mathcal{J} d\varphi, 
\end{equation}
where the Jacobian in Boozer coordinates is $\mathcal{J} = (G + \iota I)/B^2$. Since the flux-surface average is defined as
\begin{equation}
    \left\langle \cdots  \right\rangle  = \frac{1}{V'(\psi)} \int_0^{2 \pi} d\theta \int_0^{2 \pi}  ( \cdots ) \mathcal{J} d\varphi, 
\end{equation}
it follows that $\langle B^2 \rangle = 4 \pi^2 (G + \iota I) / V'$ and consequently 
\begin{equation}
    \frac{V''}{V'} = - \frac{d \ln \langle B^2 \rangle}{d \psi} + \frac{d}{d \psi} \ln ( G + \iota I),
    \label{eq:V''}
\end{equation}
where the last term vanishes in a vacuum field, where $G'(psi)$ and $I\psi)$ vanish. The magnetic well is sometimes defined without this term \citep{freidberg2014}. Thus we see, as argued in Section~\ref{sec:intro}, that increasing the radial derivative of $B$ is directly related to MHD stability. Indeed, \cite[p.~24]{helander2014theory} showed that there is a mathematical relation between maximum-$J$ and magnetic-well criteria, but they are not identical. In the limit of large mirror ratio, where most particles are magnetically trapped, then the maximum-$J$ property rigorously implies a magnetic well. It is therefore not surprising that the particular configuration optimised here possesses a vacuum magnetic well.


\section{Discussion and conclusions}
Before concluding the paper, we comment on the  connection between the maximum-$J$ condition and the suppression of trapped-particle instabilities \citep{proll2012resilience,helander2013collisionless} as well as the general reduction of curvature-driven instabilities and turbulence \citep{proll2022turbulence}. These phenomena are sensitive to the relative size of the drift frequency $\omega_d = k_\alpha \omega_\alpha$ and the diamagnetic frequency
\begin{equation}
    \omega_\star = \frac{k_\alpha T}{q} \frac{ \ln n}{d \psi},
\end{equation}
where $T$ denotes the plasma temperature in units of energy, $n$ the density, and the wave vector has been taken to be ${\bf k} = k_\alpha \nabla \alpha$. It is thus of interest to assess the magnitude of the ratio
\begin{equation}
    \frac{\omega_d}{\omega_\ast} = \frac{mv^2}{4T} \frac{\hat{\omega}_\alpha}{\psi_a d\ln n/d\psi},
\end{equation}
which for a density profile $n = n_0(1-\psi/\psi_a)$ becomes $\omega_d / \omega_\ast = -\hat{\omega}_\alpha/2$ for particles moving at the thermal speed, $v = \sqrt{2T/m}$. In the QI cases studied numerically in the present paper, $\hat{\omega}_\alpha$ is of order $1/10$, making $\omega_d / \omega_\ast \ll 1$ for most of the particles in a thermal distribution. This circumstance is stabilising to trapped-electron modes even if the absence of the maximum-$J$ property. Relatively few electrons are able to satisfy the resonance condition in the limit $\omega_d/\omega_\ast \rightarrow 0$ \citep{Connor_2006}, and the nonlinear energy available to trapped electrons for driving instabilities is small \citep{mackenbach2022available,mackenbach_proll_wakelkamp_helander_2023}. The latter was recently studied and labelled as the `strong' regime \citep{rodriguez2023precession}, in which the available energy is only supplied by a small fraction of the trapped population $\sim\omega_\alpha/\omega_\star$. From a practical point of view, it is true that as long as $|\hat \omega_\alpha | \ll 1$ the magnetic field may not need to strictly satisfy the maximum-$J$ condition in order to substantially improve trapped-electron-mode stability. This condition is nevertheless helpful for MHD stability and fast-particle confinement, since particles with $\omega_\alpha = 0$ are prone collisionless losses \citep{velasco2021model, paul2022energetic}. In any case, whenever the magnetic field of any concrete stellarator is optimised, a compromise will need to find between the maximum-$J$ condition and other desirable properties.  
\par
It is fortunate, though not merely a happy coincidence, that the maximum-$J$ property is related to both MHD stability, trapped-electron-mode suppression, and good fast-ion confinement: three very beneficial properties that are otherwise considered separate and independent of each other. Quasi-symmetric stellarators struggle to possess the maximum-$J$ property, but we show that quasi-isodynamic ones are more amenable to it, and can be designed accordingly. Physically, this property is realised if all trapped particles experience average favourable magnetic curvature, and the precession frequency of these particles then has the opposite sign from the diamagnetic frequency. The behaviour of the most deeply and the most shallowly trapped particle orbits can be examined analytically. It is found that the former usually precess in the wrong (i.e. diamagnetic) direction. This behaviour may be modified by appropriate twisting and flattening of the field near the bottom of the trapping well. Barely trapped particles always precess as desired, provided the magnetic field is perfectly omnigenous. Nevertheless, omnigenity always fails for some of these particles, making them, too, precess in the wrong direction (as explicitly shown). As we have seen, despite this formal limitation, it is possible to design a vacuum QI field in which the vast majority of all particles satisfy the maximum-$J$ condition. 

\section*{Data availability}
The data that support the findings of this study are openly available at the Zenodo repository with DOI/URL 10.5281/zenodo.10203417.

\section*{Acknowledgements}
Many of the numerical utilities to handle \texttt{VMEC} equilibria \citep{hirshman1983} are based on the stellarator package \texttt{simsopt} \citep{landreman2021simsopt}. We gratefully acknowledge fruitful discussion with J. L. Velasco and R. J. J. Mackenbach. E. R. was supported by a grant by Alexander-von-Humboldt-Stiftung, Bonn, Germany, through a postdoctoral research fellowship, and A. G. supported by a grant from the Simons Foundation (560651, PH). This work has been carried out within the framework of the EUROfusion Consortium, funded by the European Union via the Euratom Research and Training Programme (Grant Agreement No 101052200 — EUROfusion). Views and opinions expressed are however those of the authors only and do not necessarily reflect those of the European Union or the European Commission. Neither the European Union nor the European Commission can be held responsible for them.  

\appendix
\section{Precession frequency}\label{app:precession}

\subsection{General expressions} 

In accordance with the discussion in Section~\ref{sec:intro}, we first turn our attention to the derivatives of the adiabatic invariant $\mathcal{J}_\parallel$. In Boozer coordinates, the magnetic field can be written as \citep{boozer1981plasma,helander2014theory}
    \begin{equation}
    {\bf B} = G(\psi) \nabla \varphi + I(\psi) \nabla \theta + B_\psi(\psi,\theta,\varphi) \nabla \psi, \label{eqn:Bcov}
    \end{equation}
the arc length element is $d\ell = (G+\iota I) d \varphi / B $, so the adiabatic invariant becomes
    $$ \mathcal{J}_\parallel = mv(G + \iota I) \int \frac{\sqrt{1-\lambda B}}{B} d\varphi, $$
and its derivatives are
    \begin{equation}
    \left.\frac{\partial \mathcal{J}_\parallel}{\partial\alpha}\right|_{\psi,\lambda} = -mv (G + \iota I) \int \left(\frac{\partial B}{\partial \alpha}\right)_{\psi,\varphi} F(\lambda, B) d\varphi,
    \label{d1}
    \end{equation} 
    \begin{equation} 
    \left.\frac{\partial \mathcal{J}_\parallel}{\partial\psi}\right|_{\alpha,\lambda} = mv \left[(G + \iota I)' \int \frac{\sqrt{1-\lambda B}}{B} d\varphi
    - (G + \iota I) \int \left(\frac{\partial B}{\partial \psi}\right)_{\alpha,\varphi} F(\lambda, B) d\varphi \right], 
    \label{d2}
    \end{equation}
with
    $$ F(\lambda,B) = \frac{1-\lambda B/2}{B^2 \sqrt{1-\lambda B}},$$
and the contribution from the boundary terms vanishing at the end points of the bounce integral.
Equation (\ref{d2}) can be simplified by using the MHD force balance equation. In particular, using Eq.~(\ref{eqn:Bcov}) to find the plasma current, computing $\mathbf{j}\times\mathbf{B}$ using the Boozer Jacobian and flux-surface averaging the $\nabla\psi$ component of the equation \citep{Kruskal-Kulsrud,helander2014theory} 
    $$ G' + \iota I' = -\mu_0p' \frac{G+\iota I}{\langle B^2\rangle}, $$
where $p(\psi)$ denotes the plasma pressure and angular brackets indicate the usual flux-surface average. Hence
    \begin{equation} 
    \left.\frac{\partial \mathcal{J}_\parallel}{\partial \psi}\right|_{\alpha,\lambda} = mv (G + \iota I) \left[\left(\frac{\iota' I }{G + \iota I}
    - \frac{\mu_0 p'}{\langle B^2 \rangle}  \right) 
    \int \frac{\sqrt{1-\lambda B}}{B} d\varphi
    - \int \left(\frac{\partial B}{\partial \psi}\right)_{\alpha,\varphi} F(\lambda, B) d\varphi \right].
    \label{eq:dJdpsi}
    \end{equation}
The bounce time is
    $$ \Delta t = \int \frac{d\ell}{v_\|} = \frac{G + \iota I}{v} \int \frac{d\varphi}{B \sqrt{1-\lambda B}}, $$
and the precession frequency $\omega_\alpha = \Delta \alpha / \Delta t$, using Eq.~(\ref{Delta alpha}), thus becomes
    \begin{equation}
     \omega_\alpha = \frac{mv^2}{q} \left[ \int  \left(\frac{\partial B}{\partial \psi}\right)_{\alpha, \varphi} 
    F(\lambda,B) \; d\varphi
    - \left( \frac{\iota' I }{G + \iota I} - \frac{\mu_0 p'}{\langle B^2 \rangle}\right) 
    \int \frac{\sqrt{1-\lambda B}}{B} d \varphi   \right]
      \bigg\slash \int \frac{d \varphi}{B \sqrt{1-\lambda B}}.
    \label{omegaa}
    \end{equation}
The same expressions may be reached by considering the bounce integral of the poloidal and radial drifts directly.
\par
\subsection{Precession for deeply-trapped particles}
The expression in Eq.~(\ref{omegaa}) is valid in any stellarator with flux surfaces, but becomes much simpler for deeply trapped particles. In this case $1-\lambda B \ll 1$ in the full integration domain, which also shrinks as $\lambda^{-1}$ approaches the minimum of $|\mathbf{B}|$. This makes the first term in the square brackets dominate over the second one, and $F(\lambda,B) \approx 1/(2B^2 \sqrt{1-\lambda B})$ independently of the shape of the magnetic well. To show this, one can approximate the bottom of the well as $B\approx B_0(1+B^{(n)}\varphi^n/n!)$ and evaluate for a deeply trapped particle with bounce point $\varphi_b$,
\begin{subequations}
    \begin{gather}
        \int_0^{\varphi_b}\frac{\sqrt{1-\lambda B}}{B}\mathrm{d}\varphi\approx \frac{\sqrt{\pi}}{2B_0}\frac{\Gamma(1+1/n)}{\Gamma(3/2+1/n)}\left(\lambda B_0\frac{B^{(n)}}{n!}\right)^{-1/n}(1-\lambda B_0)^{1/2+1/n}, \\
        \int_0^{\varphi_b}\frac{1}{\sqrt{1-\lambda B}}\mathrm{d}\varphi\approx \sqrt{\pi}\frac{\Gamma(1+1/n)}{\Gamma(1/2+1/n)}\left(\lambda B_0\frac{B^{(n)}}{n!}\right)^{-1/n}(1-\lambda B_0)^{-1/2+1/n}.
    \end{gather} \label{eqn:bot_well_walpha_p_vanish}
\end{subequations}
The second term in the square bracket of Eq.~(\ref{omegaa}) thus yields a contribution that is $\sim(1-\lambda B_0)$ smaller than the first, and thus vanishes for deeply trapped particles regardless of the shape of the well. 

As a result, at the bottom of the magnetic well (i.e., at the point where ${\bf B} \cdot \nabla B = 0$) the precession frequency becomes simply
\begin{equation}
     \omega_\alpha = \frac{mv^2}{2qB} \left(\frac{\partial B}{\partial \psi}\right)_{\alpha, \varphi}.
    \label{omegaa-deeply-trapped}
\end{equation}
This result regarding the local role of $\partial_\psi B$ also applies to the mostly shallowly trapped particles, which spend most of the time close to the turning point where the magnetic field strength reaches its maximum on the field line in question, $\lambda B_{\rm max} \simeq 1$. 
\par

The partial derivative of $B$ keeping $(\alpha,\varphi)$ constant can be expressed in terms of common Boozer derivatives at fixed $(\theta,\varphi)$ by using the chain rule,
\begin{equation}
    \partial_\psi B|_{\alpha,\varphi}=\partial_\psi B|_{\theta,\varphi}+\iota'\varphi\partial_\theta B|_{\psi,\varphi}. \label{eqn:dBdpsi}
\end{equation}
In this expression, the second term must vanish at the bottom of the magnetic well if the field is QI (in fact \textit{pseudosymmetry} is sufficient \citep{mikhailov2002,skovoroda2005}), as can be seen from the following argument. By definition, $(\iota \partial_\theta+\partial_\varphi)B=0$ at the bottom of the magnetic well. Moreover, in a QI field, the level contours of $B = |{\bf B}|$ are poloidally closed, which implies that $(c\partial_\theta+\partial_\varphi)B=0$ for some $c(\psi,\theta,\varphi)\ne 0$ that cannot equal $\iota$, at least not at all minima of $B$. At the bottom of the well, then, the directional derivative of $B$ vanishes in two independent directions tangential to the magnetic surface, which implies that it vanishes in any such direction including $\partial_\theta B = 0$. As a result, Eq.~(\ref{eqn:dBdpsi}) becomes 
\begin{equation}
    \partial_\psi B|_{\alpha,\varphi}=\partial_\psi B|_{\theta,\varphi}. 
\end{equation}
A similar argument also applies to maxima of the field strength.

\subsection{Near-axis expansion}

We now consider the region close to the magnetic axis, expanding in the minor-radius variable $r = \sqrt{2 \psi / \bar B}$, where $\bar B$ is some reference magnetic-field strength. To second order, this expansion is \citep{garrenboozer1991a,rodriguez2023higher}
    $$ B = B_0(\varphi) + r B_1(\alpha,\varphi) + r^2 B_2(\alpha,\varphi), $$
with
\begin{subequations}
\begin{gather}
    B_1(\alpha,\varphi) = B_0(\varphi) d(\varphi) \cos\left[\alpha + \nu(\varphi) \right], \\
    B_2(\alpha,\varphi) = B_{20}(\varphi) 
    + B_{2c}(\varphi) \cos 2 \left[\alpha + \nu(\varphi) \right]
    + B_{2s}(\varphi) \sin 2 \left[\alpha + \nu(\varphi) \right].
\end{gather} \label{eqn:define_B_QI_nae}
\end{subequations}
To check for omnigeneity, let us write the integral in Eq.~(\ref{d1}) for $\partial_\alpha\mathcal{J}_\parallel$,
\begin{equation} 
    \int \frac{\partial B}{\partial \alpha} F(\lambda, B) d\varphi \approx 
    \int \left( r \frac{\partial B_1}{\partial \alpha} + r^2 \frac{\partial B_2}{\partial \alpha} \right)
    \left[ F(\lambda,B_0) + r B_1 \frac{\partial F}{\partial B} \right] d \varphi, 
    \label{1st integral}
\end{equation}
where $\partial F / \partial B = (\partial F(\lambda,B_0)) / \partial \varphi) / B_0'(\varphi)$ and the integral is meant between bouncing points $\varphi_{R,L}$. Expanding to second order in $r$ and integrating by parts\footnote{Some care is called for when computing perturbations of bounce integrals since the limits of integration (bounce points) are themselves perturbed and one must guarantee that the argument in $\sqrt{1-\lambda B}$ is positive. This can be nevertheless done carefully to arrive at the same final QI conditions.}
\begin{multline}
    \int F(\lambda,B_0) \left[ r \frac{\partial B_1}{\partial \alpha} + r^2 \frac{\partial B_2}{\partial \alpha} 
    - r^2 \frac{\partial}{\partial \varphi} \left( \frac{B_1}{B_0'} \frac{\partial B_1}{\partial \alpha} \right) \right] d\varphi \\
     = \int F(\lambda,B_0) \left[ - r d B_0 \sin(\alpha + \nu) 
    + 2 r^2 \left( B_{2s} \cos 2 (\alpha + \nu) - B_{2c} \sin 2 (\alpha + \nu)\right) \right.
    \\
     \left. + r^2 \frac{\partial}{\partial \varphi} \left( \frac{B_0^2 d^2}{2B_0'} \sin 2(\alpha + \nu)\right) \right] d\varphi,  
    \label{2nd-order omnigenity}
\end{multline}
where the integral is taken between bounce points of $B_0(\varphi)$. If the field is to be omnigenous, $\partial_\alpha \mathcal{J}_\parallel = 0$, this quantity must vanish for all $\lambda \in [B_{\rm max}^{-1},B_{\rm min}^{-1}]$ and all $\alpha \in [0,2\pi]$. 
For a stellarator-symmetric field, $B(r,\alpha,\varphi) = B(r,-\alpha,-\varphi)$, we have the relations
\begin{subequations}
\begin{gather}
    B_0(\varphi) = B_0(-\varphi), \\
    d(\varphi) \cos[\alpha + \nu(\varphi) ] = d(-\varphi) \cos[-\alpha + \nu(-\varphi)],  
\end{gather}\label{eqn:omn_cond_1st}
which are satisfied if $d$ is chosen to be odd in $\varphi$, and $\nu = (n+1/2) \pi$. As a result of these constraints, the first-order contribution to Eq.~(\ref{2nd-order omnigenity}) in $r$ vanishes. That at 2nd-order will also vanish iff
\begin{gather}
    B_{2s}(\varphi) = -B_{2s}(-\varphi), \\
    B_{2c}(\varphi) = \frac{\partial}{\partial \varphi} \left( \frac{B_0^2 d^2}{4B_0'} \right).  \label{eqn:B2c-condition}
\end{gather}
\end{subequations}
These relations form the basic requirements for omnigenity close to the magnetic axis and were recently derived by \cite{rodriguez2023higher} by a different method that avoids the explicit integrals. We refer to that paper for further details and discussion.
\par
Employing the same near-axis explansion to (\ref{eq:dJdpsi}) gives in analogy to  Eq.~(\ref{2nd-order omnigenity}) for a general stellarator,
\begin{multline}
    \int F(\lambda,B)\left(\frac{\partial B}{\partial\psi}\right)_{\alpha,\varphi}\mathrm{d}\varphi= \\
    \frac{2}{\bar{B}}\left[\frac{1}{2r}\int F(\lambda,B_0)B_1\mathrm{d}\varphi+\int F(\lambda,B_0)\left[B_2-\frac{\partial}{\partial\varphi}\left(\frac{B_1^2}{2B_0'}\right)\right]\mathrm{d}\varphi\right], \label{eqn:integ_2}
\end{multline}
where the integrals are, once again, taken between bounce points of $B_0(\varphi)$. Substituting the omnigeneous forms of the magnetic field functions into this integral, and noting that the enclosed toroidal current $I\sim r^2$ (in the absence of an infinite current density on axis), one obtains $\omega_\alpha=\omega_{\alpha,\mathrm{vac}}+\omega_{\alpha,p}$ with
 \begin{equation}
        \omega_{\alpha,\mathrm{vac}} = \frac{2mv^2}{q \bar{B}}\frac{\int_{\varphi_L}^{\varphi_R}\frac{1-\lambda B_0/2}{B_0\sqrt{1-\lambda B_0}}\left[\frac{B_{20}}{B_0}-\frac{1}{4B_0}\left(\frac{B_0^2d^2}{B_0'}\right)'\right]\mathrm{d}\varphi}{\int_{\varphi_L}^{\varphi_R}\frac{\mathrm{d}\varphi}{B_0\sqrt{1-\lambda B_0}}}, \label{eqn:nae_w_alpha}
    \end{equation}
     \begin{equation}
        \omega_{\alpha,p} = \frac{2mv^2}{q \bar{B}}\frac{\mu_0p_2}{\langle B_0^2\rangle}\frac{\int_{\varphi_L}^{\varphi_R}\frac{\sqrt{1-\lambda B_0}}{B_0}\mathrm{d}\varphi}{\int_{\varphi_L}^{\varphi_R}\frac{\mathrm{d}\varphi}{B_0\sqrt{1-\lambda B_0}}}, \label{eqn:nae_w_alphap}
    \end{equation}
where $p(r) = p_0 + p_2 r^2 + \cdots$ and ${\varphi_L}$ and $\varphi_R$ denote the toroidal locations of the bounce point to the left (L) and right (R) of the bottom of the well (for stellarator symmetry $\varphi_L=-\varphi_R$). Although we have used the notation $\omega_{\alpha,\mathrm{vac}}$ and $\omega_{\alpha,p}$, one should not forget that the former includes pressure effects through $B_{20}$. The flux-average of $B_0^2$ appearing in this expression is equal to
     \begin{equation} \langle B_0^2\rangle = 2 \pi \bigg\slash \int_0^{2\pi} \frac{\mathrm{d}\varphi}{B_0^2}.
     \end{equation}

\subsection{Omnigeneity breaking} \label{app:omnigeneity-breaking-wa}
Because the expression in Eq.~(\ref{omegaa}) is valid for any stellarator, it provides an opportunity to learn how deviations from exact omnigenity affect the precession frequency $\omega_\alpha$. We consider this question from the near-axis perspective.
\par
\subsubsection{First order breaking}
Let us start with the QI requirement at first order in $r$. As we saw following Eq.~(\ref{2nd-order omnigenity}), the condition of omnigeneity requires the functions $d$ and $\nu$ to be odd and even, respectively, in $\varphi$ (see \cite{plunk2018,camacho-mata-2022,rodriguez2023higher}). This symmetry is crucial in a quasi-isodynamic stellarator, as it causes the radial drift on either side of the magnetic well to cancel to leading order (as discussed in the main text). When the condition of omnigeneity is satisfied exactly, the $1/r$ contribution to $\omega_\alpha$ in Eq.~(\ref{eqn:integ_2}) vanishes. This balance is delicate, however, as any small deviation can lead to a large precession, restoring the $O(1/r)$ dependence. This behaviour can be understood as the result of a finite poloidal drift occurring over an increasingly smaller flux surface as the magnetic axis is approached. The contribution may be  written as
\begin{equation}
    \omega_{\alpha,-1}^{\mathrm{non-QI}}=\frac{mv^2}{q \bar{B}}\frac{1}{r}\left[{\int_{\varphi_L}^{\varphi_R}\frac{\mathrm{d}\varphi}{B_0\sqrt{1-\lambda B_0}}}\right]^{-1}\int_{\varphi_L}^{\varphi_R} F(\lambda,B_0)B_1(\alpha,\varphi)\mathrm{d}\varphi \label{eqn:qi-1/r-wa}
\end{equation}
following directly from Eq.~(\ref{eqn:integ_2}). 
\par
This term is in principle always present since omnigeneity can never be realised exactly except in trivial cases \citep{Cary1997,landreman2012,plunk2019direct, rodriguez2023higher}. In any non-axisymmetric torus, the requirements of omnigeneity clash with the required periodicity of the solution in $\theta$ and $\varphi$. Formally, periodicity requires $\nu(-\pi)=\nu(\pi)+2\pi(\iota-N)$ in our near-axis description, which is incompatible with its even symmetry. This condition \textit{is necessarily broken} at the points of maximum field strength, which we may write $\nu=\pi/2+\Tilde{\nu}$ and $\tilde{\nu}$ is odd in $\varphi$.Barely trapped particles will always exerience this lack of omnigeneity. 
\par
Using this notation, we may rewrite the $O(1/r)$ contribution to the precession, 
\begin{equation}
    \omega_{\alpha,-1}^{\mathrm{non-QI}}=-\frac{mv^2}{q \bar{B}}\frac{\cos\alpha}{r}\left[{\int_{\varphi_L}^{\varphi_R}\frac{\mathrm{d}\varphi}{B_0\sqrt{1-\lambda B_0}}}\right]^{-1}\int_{\varphi_L}^{\varphi_R} F(\lambda,B_0)B_0 d \sin\tilde{\nu}\mathrm{d}\varphi, \label{eqn:qi-1/r-wa}
\end{equation}
to which contributions arise from any particles that venture into the region where omnigeneity is broken, i.e. $\sin\tilde{\nu}\neq0$. For these orbits, the non-omnigeneous part of the precession will always dominate for sufficiently small $r$. Because of the $\cos\alpha$ dependence of the precession, these particles will precess in either poloidal direction depending on the field line considered. The immediate implication of this behaviour is that in a \textit{realistic} magnetic field (one that allows for the \textit{necessary} breaking of QI), the maximum-$J$ property is technically not possible close to the magnetic axis, independently of plasma pressure. 
\par
This somewhat bleak result appears to rule out maximum-$J$ behaviour in QI configurations, much as in quasisymmetric stellarators or tokamaks. It should however be noted that the factors that determine the size of the $1/r$ terms are naturally small. At the top of the magnetic well, where the conditions of exact omnigeneity need to be relaxed, it is by construction the case that the function $d$ vanishes. That is, the magnetic field perturbation $B_1$ is small in the vicinity of the maximum, as is necessary for the consistency with a locally straight magnetic field (and for the contour of $B_\mathrm{max}$ to be independent of the poloidal angle). By shrinking the region in which omnigeneity is broken, the number of particles for which the precession frequency diverges as $1/r$ can be made arbitrarily small\footnote{There is however a price to be paid in terms of geometric complexity, see \cite{camacho-mata-2022}.}. Only extremely close to the axis is the divergent behaviour of the precession frequency significant.
\par
The size of this region can be gauged by estimating the minor radius $r_a$ below which the $1/r$ behaviour, Eq.~(\ref{eqn:qi-1/r-wa}), dominates over the second-order contributions, Eq.~(\ref{eqn:nae_w_alpha}). By way of example, we define and compute the critical radius  $r_a=\sqrt{\sum_k w_{\alpha,-1}^2/\sum_k w_{\alpha,0}^2}$ (with the $-1$ and 0 indices referring to the respective $r$ scalings) for the near-axis fields in Fig.~\ref{fig:qi_analysis}. The amplification of the precession and its variation from field line to field line is especially noticeable in Fig.~\ref{fig:qi_analysis}b for which $r_a\sim0.003$, driven by the presence of `buffer' regions \citep{plunk2019direct,camacho-mata-2022}. It may perhaps appear counterintuitive that close to the axis, where the field is most omngninous, the variation between field lines (an indicator of departure from omnigeneity) increases. However, the bounce-averaged radial drift $\omega_\psi$ {\em does} decrease linearly in $r$ as expected from Eq.~(\ref{2nd-order omnigenity})), indicating that omngeneity {\em does} become perfect in the limit $r \rightarrow 0$. The existence of a region very close to the axis in which the maximum-$J$ condition cannot be satisfied is nevertheless unavoidable, but this region can in principle be made arbitrarily small.
\par
\subsubsection{Second order breaking}
From Fig.~\ref{fig:qi_analysis} it is clear that there are non-QI contributions even if the leading first order field is precisely QI. In the exactly omnigenous limit, the contribution to $\omega_\alpha$ from second-order terms is $\alpha$-independent, but one regains some $\alpha$ dependence when omnigeneity is broken. Taking into account the symmetry conditions on the various functions that constitute the near-axis description, we may write the non-QI contribution to the precession as
\begin{multline}
    \omega_{\alpha,0}^\mathrm{non-QI} = -\frac{2mv^2}{q \bar{B}}\left[{\int_{\varphi_L}^{\varphi_R}\frac{\mathrm{d}\varphi}{B_0\sqrt{1-\lambda B_0}}}\right]^{-1}\cos2\alpha\times\\
    \times\int_{\varphi_L}^{\varphi_R}F(\lambda,B_0)\left\{\left[B_{2c}-\frac{1}{4}\left(\frac{B_0^2d^2}{B_0'}\right)'\right]\cos 2\tilde{\nu}+\left[B_{2s}+\frac{B_0^2d^2}{2B_0'}\Tilde{\nu}'\right]\sin2\tilde{\nu}\right\}\mathrm{d}\varphi. \label{eqn:2nd-order-deviation-wa}
\end{multline}
Two limits of this expression are of particular interest. The first limit concerns deeply trapped particles, which are relatively easily well confined and may therefore be omnigenous to second order. Then $\tilde \nu = 0$ and 
\begin{equation}
    \omega_{\alpha,0}^\mathrm{non-QI}=-\frac{2mv^2}{qB_0^2}\left[B_{2c}-\frac{1}{4}\left(\frac{B_0^2d^2}{B_0'}\right)'\right]\cos 2\alpha. \label{eqn:wa-non-qi-deep}
\end{equation}
The second limit of interest is realised when $B_{2c}$ has its QI form, Eq.~(\ref{eqn:B2c-condition}). Then the first term in square brackets of the second integral vanishes, leaving a contribution that vanishes unless omnigeneity is broken at first order. The QI contribution $\omega_{\alpha,\mathrm{vac}}$ in Eq.~(\ref{eqn:nae_w_alpha}) remains unchanged. 

\subsection{Stellarator-symmetry breaking}
The expressions above were derived under the assumption of stellarator symmetry. This is a convenient simplifying assumption, under which functions acquire a definite parity in $\varphi$, which simplifies bounce integrals. If $B_0$ is not symmetric in $\varphi$, the expressions need to be revisited although most of the considerations for symmetric fields continue to hold. 
\par
To deal with the stellarator asymmetric case, we employ the notation of \cite{rodriguez2023higher} and denote the `bounce mapping' by $\eta_0(\varphi)$. This is the function that maps a point on one side of the magnetic well defined by $B_0$ to the other side of the well, i.e., $B_0(\varphi) = B_0[\eta_0(\varphi)]$. Symmetry conditions on the functions $B_0$, $d$ etc. then apply with respect to the mapping $\eta_0$ \citep{rodriguez2023higher}. 
\par
With this in mind, we proceed from Eq.~(\ref{eqn:integ_2}) without making any assumption of symmetry.\footnote{It can be rigurously shown that such an expression holds even upon breaking of stellarator symmetry to the order shown.} The first term, corresponding to the $1/r$ order, vanished due to the QI condition and the odd symmetry of $B_1$. It can be shown that this vanishing holds even when stellarator symmetry is abandoned, 
\begin{equation}
    \int_{\varphi_{L}}^{\varphi_{R}} F(\lambda,B_0)B_1(\varphi)\mathrm{d}\varphi=\int_{\varphi_{R}}^{\varphi_{L}} F(\lambda,B_0)\underbrace{B_1[\eta_0(\bar{\varphi})]\eta_0'(\bar{\varphi})}_{\stackrel{\mathrm{QI}}{=}B_1(\bar{\varphi})}\mathrm{d}\bar{\varphi}=0, \label{eqn:1st-order-SS-int}
\end{equation}
where we changed variables $\varphi=\eta_0(\bar{\varphi})$ for the first equality, applied the first-order QI condition in a generally non-symmetric QI field [Eq.~(19) in \cite{rodriguez2023higher}], and noted that the second integral is exactly the negative of the first one. This proves that the QI property is sufficient to make the integral vanish, without any stellarator symmetry requirement being necessary. This should not come as a surprise.
\par
The QI condition at second order may be succinctly written as an equation for the $\alpha$-dependent part of $B_2$ as (Eq.~(C12) in \cite{rodriguez2023higher}),
\begin{equation}
B_2(\theta,\varphi)-\bar{B}_2(\varphi)-\partial_\varphi\left(\frac{B_1^2}{2B_0'}\right)=\eta'_0\left[B_2(\theta,\varphi)-\bar{B}_2(\varphi)-\partial_\varphi\left(\frac{B_1^2}{2B_0'}\right)\right]_b, \label{eqn:B2-qi-condition}
\end{equation}
where the subscript $b$ denotes the evaluation of the expression inside the brackets at $\eta_0$ and the barred $\bar{B}_2$ represents the poloidal average of $B_2$. Following the same substitution trick as for Eq.~(\ref{eqn:1st-order-SS-int}), the $\theta$ dependent part of the order $O(1)$ contribution to $\omega_\alpha$ can be shown to vanish, only leaving the $\theta$ (or $\alpha$) dependent piece. The conclusionis that the expression for the omnigeneous part of $\omega_{\alpha,0}$ is the same whether or not the field is stellarator symmetric, with the difference that in the asymmetric case usually $\varphi_L\neq-\varphi_R$. In addition, symmetry breaking will also affect the value of $B_{20}$.

\section{Radial derivative of $B$ for deeply trapped particles in QI configurations} \label{sec:appB20Deep}
As we have seen in the main text, the average radial derivative of $B$ at the minimum along the field, $B_{20}=\Bar{B}\partial_\psi B/2|_{\psi=0}$, is key in determining the tangential drift of trapped particles. This term is related to the magnetic well (\ref{eq:V''}), which plays an important role for MHD stability \citep{freidberg2014,landreman2020magnetic, rodriguez2023mhd}.  
\par
Regardless of whether the field is QI, the term $B_{20}$ must have a form that is consistent with the equilibrium and solenoidal properties of the magnetic field. That is, it must be consistent with $\nabla\cdot\mathbf{B}=0$ and $\mathbf{j}\times\mathbf{B}=\nabla p$. Within the asymptotic framework of the near-axis description of the field, the function $B_{20}$ is only partially constrained by the properties of the magnetic field on the axis and to first order. There is in general some freedom in its choice. However, at `straight' sections where the curvature of the magnetic axis vanishes, the constraints of the equations become particularly stringent leaving no freedom. $B_{20}$ becomes uniquely determined by lower-order choices and the plasma pressure. In this Appendix we find determine its value and explore formal consequences of it.
\par
For this purpose, we focus on the minimum of $|\mathbf{B}|$ along field, where deeply trapped particles reside and $\kappa=0$. We are interested in relating the magnetic field magnitude $B_{20}$ to lower-order quantities in the near-axis expansion and to the flux-surface geometry. To do so, and as part of what is known as the \textit{inverse-coordinate} approach to the near-axis expansion \citep{garrenboozer1991a}, we describe flux surfaces in terms of Boozer coordinates (our independent set of coordinates) respect to a signed Frenet-Serret frame \citep{plunk2019direct} $\{\hat{b},\hat{\kappa},\hat{\tau}\}$ as $\mathbf{x}-\mathbf{r}_0=X\hat{\kappa}+Y\hat{\tau}+Z\hat{b}$. Here $\mathbf{r}_0$ represents the magnetic axis and $\{X,Y,Z\}$ are functions describing the flux-surface shape. The near-axis expansion framework is designed to uncover the connection between the $|\mathbf{B}|$ and these features. The relevant equations and constructions can be found in \cite{landreman2019}, to which we refer for further details. We shall invoke results from this work, reproducing the necessary expressions. In what follows, the symbol $f_{ij}$ represents the expansion coefficients of the function $f$, where the subscript $i$ corresponds to the power of $r$ in the expansion, and $j$ equals $0,~c$ or $s$ depending on the $\theta$ harmonic it represents (constant, cosine or sine). At higher order generalised notation can be devised \citep{rodriguez2021anis}.
\par
We first consider the shape of flux surfaces in the direction normal to the magnetic axis and begin with the function $X_{20}$, which describes a rigid shift of the flux surfaces relative to each other (akin to the Shafranov shift \citep{rodriguez2023mhd,landreman2021a}). This function can be read off from Eq.~(A34) of \cite{landreman2019},
\begin{multline}
    X_{20}=\frac{1}{\kappa \ell'}\left\{Z_{20}'-\frac{1}{\ell'}\left[-\frac{G_0^2 B_{20}}{B_0^3}+\frac{3G_0^2(B_{1c}^2+B_{1s}^2)}{4B_0^4}+\right.\right.\\
    \left.\left.+\frac{G_0(G_2+\iota_0 I_2)}{B_0^2}-\frac{X_{1c}^2+X_{1s}^2}{4}(\kappa \ell')^2-\frac{q_c^2+q_s^2+r_c^2+r_s^2}{4}\right]\right\}, \label{eqn:X20}
\end{multline}
where all the expressions on the right hand side except $B_{20},~G_2$ and $I_2$ are first-order quantities. 
\par
Since $\kappa$ appears in the denominator of this expression, to avoid an unphysical diverging shift of flux surfaces at the point where $\kappa=0$, $B_{20}$ is highly constrained at this point. If the curvature has a zero of order $v$, then we expect the first $v-1$ derivatives of $B_{20}$ to be determined by Eq.~(\ref{eqn:X20}). Let us momentarily focus on $B_{20}(\varphi=0)$ for $\kappa(\varphi=0)=0$. We shall assume stellarator symmetry and the configuration to be QI to first order, so that $\Bar{d}=d/\kappa$ is even in $\varphi$, the (signed) curvature is odd, the torsion even, $B_0$ even, and $\sigma$ odd ($\sigma=(B_{1s}Y_{1s}+B_{1c}Y_{1c})/\bar{B}\kappa$ is a measure of up-down asymmetry). With this in mind, using the expression for $Z_{20}$ and $q_i$ from Eqs.~(A24), (A27) and (A37-40) of \cite{landreman2019}, we find\footnote{All functions of $\varphi$ are meant to be evaluated at $\varphi=0$, but we avoid writing this explicitly for simplicity of notation.}
\begin{multline}
    \frac{B_{20}}{B_0}=-\frac{\mu_0 p_2}{B_0^2}+\frac{1}{4(\ell')^2}\left[\bar{d}\bar{d}''\left(1-\frac{\bar{B}^2}{B_0^2\Bar{d}^4}\right)+\left(\frac{\bar{B}}{B_0\bar{d}}\right)^2(\sigma')^2-\frac{\bar{B}^2B_0''}{B_0^3\bar{d}^2}\right]-\\
    -\frac{1}{4}\left[\left(\frac{\bar{B}\tau_0}{B_0\bar{d}}\right)^2+\left(\frac{\bar{B}\sigma'}{B_0\ell'\bar{d}}+\Bar{d}\tau_0\right)\right].
\end{multline}
To further simplify this expression, we eliminate $\sigma'$ in favour of the properties of the axis and first order quantities. Using the near-axis Riccati $\sigma$-equation \citep[Eq.~(A21)]{landreman2019}, assuming an exact QI magnetic field at first order $B_{1c}=-B_0 d\sin\iota_0\varphi$ and $B_{1s}=B_0 d\cos\iota_0\varphi$ \citep{plunk2019direct,rodriguez2023higher},
\begin{equation}
    \sigma' = 2\left(\frac{I_2}{\bar{B}}-\tau_0\right)\frac{G_0\bar{d}^2}{\bar{B}}.
\end{equation}
With this and the relation $\ell'=G_0/B_0$ (assuming $G_0>0$) \citep[Eq.~(A20)]{landreman2019} we obtain
\begin{equation}
    \frac{B_{20}}{B_0}=-\frac{\mu_0p_2}{B_0^2}+\frac{1}{4(\ell')^2}\sum_i\mathcal{P}_i, \label{eqn:B20-min-exp}
\end{equation}
where
\begin{subequations}
    \begin{gather}
        \mathcal{P}_{\Bar{d}\Bar{d}''}=\Bar{d}\Bar{d}''\left(1-\frac{\bar{B}^2}{B_0^2\bar{d}^4}\right), \\
        \mathcal{P}_{B_0''}=-\frac{B_0''}{B_0}\frac{\bar{B}^2}{B_0^2\bar{d}^2}, \\
        \mathcal{P}_{\tau_0^2}=(\Bar{d}\tau_0\ell')^2\left(3-\frac{\bar{B}^2}{B_0^2\bar{d}^4}\right), \\
        \mathcal{P}_{I_2}=-4(\ell')^2\bar{d}^2\frac{\tau_0I_2}{\bar{B}},
    \end{gather}  \label{eqn:Ps-B20-min}
    \end{subequations}
which are the forms used in the main text. It will be convenient to choose our reference field $\bar{B}$ to equal $B$ at the bottom of the well, $B_0$, but we keep the expressions here general. 
\par
An analogous approach to avoid divergence of $X_{2c}$ yields the consistent value of $B_{2c}$ at the minimum of the magnetic field. In this case, we need Eq.~(A36) from \cite{landreman2019},
\begin{multline}
    X_{2c} = \frac{1}{\kappa \ell'}\left\{Z_{2c}'+2\bar{\iota}_0 Z_{2s}-\frac{1}{\ell'}\left[-\frac{G_0^2B_{2c}}{B_0^3}+\frac{3G_0^2(B_{1c}^2-B_{1s}^2)}{4B_0^4}-\right.\right.\\
    \left.\left.-\frac{X_{1c}^2-X_{1s}^2}{4}(\kappa\ell')^2-\frac{q_c^2-q_s^2+r_c^2-r_s^2}{4}\right]\right\}. \label{eqn:X2c}
\end{multline}
Then, as before, we may write for $B_{2c}$,
\begin{multline}
    \frac{B_{2c}}{B_0}=-\frac{1}{4(\ell')^2}\left[\bar{d}\bar{d}''\left(1+\frac{\bar{B}^2}{\bar{d}^4B_0^2}\right)+\left(\frac{\bar{B}}{\bar{d}B_0}\right)^2\frac{B_0''}{B_0}-4(\ell')^2\bar{d}^2\frac{\tau_0I_2}{\bar{B}}+\right.\\
    \left.+(\bar{d}\tau_0\ell')^2\left(3+\frac{\bar{B}^2}{\bar{d}^4B_0^2}\right)\right], \label{eqn:B2c-deep}
\end{multline}
with all quantities evaluated at the point $\varphi=0$. For a simple zero, the equivalent for $B_{2s}$ yields that $B_{2s}(\varphi=0)=0$. This is satisfied by the parity conditions derived from stellarator symmetry, and thus bring no additional information to the table. The expressions here presented may be found derived in a Mathematica script in the Zenodo repository associated with this paper. We re-emphasise that the form of these expressions assume exact QI at first order in the near-axis expansion. For more general forms of the expressions see the comments in Appendix~\ref{sec:app_opt_nae}.

\subsection{Implications of the QI condition}
We showed in Appendix~A and learnt from \cite{rodriguez2023higher} that, in an omnigeneous field, $B_{2c}$ must have a very particular form in a stellarator symmetric field. At the same time, from the conditions of equilibrium, we have also found in Eq.~(\ref{eqn:B2c-deep}) that the value of $B_{2c}$ is constrained at the bottom of the magnetic well. Thus, in principle, there are two different constraints that must be simultaneously satisfied in a QI field.
\par
To further look into this issue, we note that the $B_{2c}$ employed in this section is somewhat different from the QI form of $B_{2c}$ employed elsewhere in the text. In fact, $B_{2c}$ here (see \cite{landreman2019}) is defined respect to the helical angle $\chi=\theta-N\varphi$, where $N$ is related to the self-linking number of the signed Frenet frame of the axis \citep{rodriguez2022phases}. In contrast, the QI version of $B_{2c}$ (which we may denote by $B_{2c}^\mathrm{QI}$) is defined in Eqs.~(\ref{eqn:define_B_QI_nae}) respect to the angle $\alpha+\nu$. Therefore, 
\begin{subequations}
\begin{gather}
    B_{2c}=B_{2c}^\mathrm{QI}\cos 2(\bar{\iota}\varphi-\nu) - B_{2s}^\mathrm{QI}\sin 2(\bar{\iota}\varphi-\nu), \\
    B_{2s}=B_{2s}^\mathrm{QI}\cos 2(\bar{\iota}\varphi-\nu) + B_{2c}^\mathrm{QI}\sin 2(\bar{\iota}\varphi-\nu),
\end{gather}
\end{subequations}
where $\bar{\iota}=\iota-N$. Now, for a stellarator symmetric QI field at the bottom of the well, $\nu=\pi/2$ and $\varphi=0$, meaning that $B_{2c}=-B_{2c}^\mathrm{QI}$. Due to the sign flip, the resulting constraint from Eq.~(\ref{eqn:B2c-condition}) becomes
\begin{multline}
    \bar{d}\bar{d}''\left(1+\frac{\bar{B}^2}{\bar{d}^4B_0^2}\right)+\left(\frac{\bar{B}}{\bar{d}B_0}\right)^2\frac{B_0''}{B_0}-4(\ell')^2\bar{d}^2\frac{\tau_0I_2}{\bar{B}}+\\
    +(\bar{d}\tau_0\ell')^2\left(3+\frac{\bar{B}^2}{\bar{d}^4B_0^2}\right)=\frac{(\ell')^2}{B_0}\left(\frac{B_0^2d^2}{B_0'}\right)_{\varphi=0}'. \label{eqn:omnnn-constr-min}
\end{multline}
The condition of omnigeneity at the bottom of the trapping well thus translates into an additional constraint on the zeroth and first order components that make up the magnetic field. In the limit of a vacuum field with no secondary minima of $|\mathbf{B}|$, it follows that $\bar{d}\bar{d}''<0$: geometrically, flux surfaces must become more elongated in the binormal direction away from the minimum. This is a geometric consequence of the QI conditions. 
\par
With this constraint at hand, we may eliminate $\bar{d}''$ altogether from the expression we have for $B_{20}$ in Eq.~(\ref{eqn:B20-min-exp}). Defining $f_0=B_{20}/B_0-(B_0^2d^2/B_0')'/4B_0=f_p+f_{B_0''}+f_{I_2}+f_{\tau_0^2}+f_\mathrm{QI}$, the resulting expression is
    \begin{subequations}
        \begin{gather}
            f_p=-\frac{\mu_0p_2}{B_0^2}, \\
            f_{B_0''}=-\frac{1}{2(l')^2}\frac{\sqrt{\bar{\alpha}}}{1+\bar{\alpha}}\frac{\bar{B}}{B_0}\frac{B_0''}{B_0}, \\
            f_{\tau_0^2} = \frac{\sqrt{\bar{\alpha}}}{1+\bar{\alpha}}\frac{\Bar{B}}{B_0}\tau_0^2, \\
            f_{I_2}=-\frac{2\sqrt{\bar{\alpha}}}{1+\bar{\alpha}}\frac{I_2}{B_0}\tau_0, \\
            f_\mathrm{QI} = -\frac{1}{2B_0}\frac{1}{1+\bar{\alpha}}\left.\left(\frac{B_0^2d^2}{B_0'}\right)'\right|_{\varphi=0},            
        \end{gather} \label{eqn:f-SS-min}
    \end{subequations}
where $\bar{\alpha}=\bar{d}^4B_0^2/\bar{B}^2$.

\subsection{Stellarator symmetry breaking}
The derivation above may be extended to non-stellarator-symmetric fields, which leads to additional terms in various equations. For simplicity, we shall assume once againa that the field is exactly omnigeneous to first order and that $\nu=\pi/2$. The latter condition is not necessary as one could add an even function $\tilde{\nu}$ while still satisfying the conditions of omnigeneity \citep{rodriguez2023higher}. In the interest of brevity, however, we only consider this simplest case. 
\par
As a result of the lack of symmetry, the expression for the magnetic well $B_{20}$ changes. Proceeding in an equivalent way to that above, it can be shown that, in the expression~(\ref{eqn:B20-min-exp}), the coefficients of $B_{20}/B_0$ become (assuming $G_0>0$)
    \begin{align}
        \mathcal{P}_{\Bar{d}\Bar{d}''} & =\Bar{d}\Bar{d}''\left(1-\frac{\bar{B}^2(1+\sigma^2)}{B_0^2\bar{d}^4}\right) & \mathcal{P}_{\Bar{d}'} & =4 \Bar{d}'\ell'\frac{\bar{B}\sigma\tau}{B_0\bar{d}} \nonumber\\
        \mathcal{P}_{B_0''} & =-\frac{B_0''}{B_0}\frac{\bar{B}^2(1+\sigma^2)}{B_0^2\bar{d}^2} & \mathcal{P}_{(\bar{d}')^2} & =2(\bar{d}')^2\frac{\bar{B}^2(1+\sigma^2)}{B_0^2\bar{d}^4} \nonumber\\
        \mathcal{P}_{\tau_0^2} & =(\Bar{d}\tau_0\ell')^2\left(3-\frac{\bar{B}^2(1+\sigma^2)}{B_0^2\bar{d}^4}\right) & \mathcal{P}_{\tau_0'} & =-2\tau_0'\ell'\frac{\bar{B}\sigma}{B_0} \nonumber \\
        \mathcal{P}_{I_2} & =-4(\ell')^2\bar{d}^2\frac{\tau_0I_2}{\bar{B}}. &
    \end{align}
The main difference to the stellarator symmetric case is the modification of terms through the breaking of up-down symmetry, $\sigma$, and the appearence of three new terms. If the field is up-down symmetric at the minimum of $|\mathbf{B}|$ ($\sigma=0$), the only new term compared with the stellarator-symmetric case is that proportional to $(\bar{d}')^2$. This term is always positive, thus indicating a potential benefit of breaking stellarator-symmetry to increase the depth of the radial well.
\par
Proceeding similarly with $B_{2c}$, we find
\begin{multline}
    \frac{B_{2c}}{B_0}=-\frac{1}{4(\ell')^2}\left[\bar{d}\bar{d}''\left(1+\frac{\bar{B}^2(1-\sigma^2)}{\bar{d}^4B_0^2}\right)+\left(\frac{\bar{B}}{\bar{d}B_0}\right)^2\frac{B_0''}{B_0}(1-\sigma^2)-4(\ell')^2\bar{d}^2\frac{\tau_0I_2}{\bar{B}}+\right.\\
    \left.+(\bar{d}\tau_0\ell')^2\left(3+\frac{\bar{B}^2(1-\sigma^2)}{\bar{d}^4B_0^2}\right)+4\bar{d}'\ell'\frac{\bar{B}}{B_0\bar{d}}\sigma\tau_0-2(\bar{d}')^2\frac{\Bar{B}^2}{B_0^2\bar{d}^4}(1-\sigma^2)-2\tau_0'\ell'\sigma\frac{\bar{B}}{B_0}\right]
\end{multline}
Now, when stellarator symmetry is broken, the condition of QI at second order changes, and the self-consistency condition for $B_{2c}(0)$ will change accordingly. That is, we must check the implications of Eq.~(\ref{eqn:B2-qi-condition}), drawn from \cite{rodriguez2023higher} (see Eq.~(26b) there). At the bottom of the well it is always true that $\eta_0'\approx -1$, by virtue of having a minimum (and thus in a local Taylor expansion in $\varphi$, a leading even power of $\varphi$). Because $d$ must vanish at this position rapidly enough to avoid the loss of confinement of deeply trapped particles (see the discussion on pseudosymmetry and the behaviour at the bottom of the well in \cite{rodriguez2023higher}), $B_{2c}(0)$ must be the same as in the stellarator symmetric case, at least under the assumption made on the form of $\nu$.
\par
This way, we form the equivalent to Eq.~(\ref{eqn:f-SS-min}) when stellarator symmetry is broken, which becomes
\begin{align}
    f_p & =-\frac{\mu_0p_2}{B_0^2}, & f_{\bar{d}'} =& 2\frac{\Bar{B}}{B_0\bar{d}\ell'}\frac{\sigma\tau_0}{1+\bar{\alpha}-\sigma^2}\bar{d}', \nonumber \\
    f_{B_0''} & =-\frac{1}{2(l')^2}\frac{\sqrt{\bar{\alpha}}}{\bar{\alpha}+1-\sigma^2}\frac{\bar{B}}{B_0}\frac{B_0''}{B_0}, & f_{(\bar{d}')^2}=& \frac{1}{1+\bar{\alpha}-\sigma^2}\left(\frac{\bar{d}'}{\ell'}\right)^2, \nonumber\\
    f_{\tau_0^2} & = \frac{\sqrt{\bar{\alpha}}}{1+\bar{\alpha}-\sigma^2}\frac{\Bar{B}}{B_0}\tau_0^2, & f_{\tau_0'} =& -\frac{\tau_0'}{\ell'}\frac{\bar{B}}{B_0}\frac{\sigma}{1+\bar{\alpha}-\sigma^2}, \nonumber\\
    f_{I_2} & =-\frac{2\sqrt{\bar{\alpha}}}{1+\bar{\alpha}-\sigma^2}\frac{I_2}{B_0}\tau_0, & f_\mathrm{QI} =& -\frac{1}{2B_0}\frac{1}{1+\bar{\alpha}-\sigma^2}\left.\left(\frac{B_0^2d^2}{B_0'}\right)'\right|_{\varphi=0}.
\end{align}
Some of these expressions are used in the main text.

\section{The magnetic mirror problem} \label{app:magMirror}
There is an obvious analogy between magnetic mirrors and quasi-isodynamic stellarators. Indeed, the latter are sometimes referred to as \textit{linked mirrrors}. This analogy is especially accurate near the maxima and minima of $|\mathbf{B}|$ along field lines, which tend to be located in places where the curvature is small (and vanishes on the magnetic axis) and where trapped particles thus have little sense of the toroidicity of the field. This analogy provides us with a practical tool to assess the behaviour of the QI field from the perspective of a straight magnetic mirror.

\subsection{Near-axis expansion procedure}
The near-axis approximation (commonly referred to as the \textit{paraxial approximation} in the context of magnetic mirrors) is common in the study of magnetic mirrors \citep{furth1964closed,kadomtsev1967plasma,catto1981generalized,savenko2006mhd}. In this appendix, we adopt this approximation within the inverse-coordinate approach used for stellarators. We do so in an attempt to simplify the comparison and to clarify the quadrupole ansatz traditionally used in the analysis of magnetic mirrors. 
\par
We focus on the case of a vacuum field and describe the field lines as curves of constant $\psi$ and $\alpha$, whose position is given by $\mathbf{x}(\psi,\alpha,z)=X\hat{\mathbf{x}}+Y\hat{\mathbf{y}}+(z+Z)\hat{\mathbf{z}}$ in Cartesian coordinates. The $z$-axis can be thought of as the `magnetic axis', about which we will perform the expansion, and the function $Z$ can without loss of generality be chosen to vanish. 
\par
In order to find the governing equations, we write the vacuum condition of the magnetic field and its Clebsch representation as \citep{d2012flux}
\begin{equation}
    \nabla\Phi=\nabla\psi\times\nabla\alpha, \label{eqn:vacuum-mirror-cocon}
\end{equation}
or in the inverse-coordinate representation $\{\psi,\alpha,z\}$,
\begin{equation}
    \partial_z\mathbf{x}=\partial_\psi\Phi \partial_\alpha\mathbf{x}\times\partial_z\mathbf{x}+\partial_\alpha\Phi\partial_z\mathbf{x}\times \partial_\psi\mathbf{x} +\partial_z\Phi \partial_\psi\mathbf{x}\times \partial_\alpha\mathbf{x}. \label{eqn:vacuum-mirror-cocon-inv}
\end{equation}
To relate it to the magnetic field magnitude, we then write what we shall refer to as the Jacobian equation,
\begin{equation}
    \left(\frac{\partial_z\Phi}{B}\right)^2=|\partial_z\mathbf{x}|^2, \label{eqn:jac-equation=-mirror}
\end{equation}
which follows from taking the scalar product of Eq.~(\ref{eqn:vacuum-mirror-cocon}) with its right-hand-side, using the dual relation $\nabla\psi\times\nabla\alpha=\mathcal{J}^{-1}\partial_z\mathbf{x}=\mathbf{B}$. 
\par
As usual, we need to expand these equations by expressing all functions as Taylor-Fourier expansions in the pseudo-radial coordinate $r=\sqrt{2\psi/\bar{B}}$ and $\alpha$. The latter is a field-line label, and because we are in a straight, non-toroidal system, we may take it as our poloidal angle. (In the absence of toroidal flux surfaces, there is no need to introduce a toroidal or a helical angle.) The relevant functions in the expansion are $B,~X$ and $Y$ as well as the scalar potential $\Phi$. We shall use the same index notation as in the regular stellarator near-axis expansion, so that, for instance, 
\begin{multline}
    X(\psi,\alpha,z) = r \left[ X_{1s}(z) \sin \alpha + X_{1c}(z) \cos \alpha  \right] 
    + \\
    +r^2 \left[ X_{20}(z) + X_{2s}(z) \sin 2\alpha + X_{2c}(z) \cos 2 \alpha  \right] 
    + \cdots
\end{multline}
Here, the functions $X_{1c}(z)$ and $X_{1s}(z)$ are considered to be first-order inputs to the problem alongside $B_0(z)$, the magnetic field strength on axis. 

\subsubsection{Zeroth order (\ref{eqn:jac-equation=-mirror})}
To leading order $O(r^0)$ of Eq.~(\ref{eqn:jac-equation=-mirror}),
\begin{equation}
    \Phi_0=\int B_0(z)\mathrm{d}z,
\end{equation}
where $B_0$ is taken to be positive. 

\subsubsection{Leading order (\ref{eqn:vacuum-mirror-cocon-inv})}
To leading order, $O(r^{-1})$, there are three different equations we obtain from the three projections along the Cartesian basis of Eq.~(\ref{eqn:vacuum-mirror-cocon-inv}). From the $\hat{z}$-projection we have
\begin{equation}
    X_{1c}Y_{1s}-X_{1s}Y_{1c}=\frac{\bar{B}}{B_0(z)}, \label{eqn:mirror-expansion-flux}
\end{equation} 
which is a statement of magnetic flux conservation along the flux tube labelled by $r$. This is of course fully analogous to the situation in a stellarator, as it is a property inherent to any flux tube.
\par
For a non-vanishing magnetic field $B_0\neq0$, the other two components of Eq.~(\ref{eqn:vacuum-mirror-cocon-inv}) to $O(r)$ can be combined linearly using $X_{1c}$ and $X_{1s}$. The result is that $\Phi_{1,c} = \Phi_{1,s}= 0$. There is, in other words, no first-order contribution to the vacuum magnetic potential. 
\par
Next, the first-order $O(r)$ form of Eq.~(\ref{eqn:jac-equation=-mirror}) yields the natural conclusion that $B_{1s}=0$ and $B_{1c}=0$, so that there is no $\alpha$-variation of the magnetic field strength to first order. This is an expected result, as we well know that, in the absence of a pressure gradient, the perpendicular gradient of $|\mathbf{B}|$ is proportional to the curvature vector. In the vicinity of a straight axis, there is therefore no first-order variation of the magnetic field strength. This circumstance sets the straight mirror apart from a typical QI stellarator, in which the toroidal nature of the field forces the field to exhibit such a variation in most places along the magnetic axis. The leading form of the field is then that at next order, which must be quadrupole-like.

\subsubsection{Zeroth order (\ref{eqn:vacuum-mirror-cocon-inv})}
Let us now turn to the next order of Eq.~(\ref{eqn:vacuum-mirror-cocon-inv}), which is $O(r^0)$. At this order we have a total of six independent equations, as each of the projections has $\cos\alpha$ and $\sin\alpha$ terms.
\par
From the $\hat{z}$ projection, and constructing appropriate linear combinations of the equations, we may write the following equations for $Y_{2c}$ and $Y_{2s}$,
\begin{subequations}
    \begin{multline}
        Y_{2c}=\frac{X_{1c}^2-X_{1s}^2}{X_{1c}^2+X_{1s}^2}Y_{20}+\frac{X_{1c}Y_{1s}-X_{1s}Y_{1c}}{X_{1c}^2+X_{1s}^2}X_{2s}+\frac{X_{1c}Y_{1c}+X_{1s}Y_{1s}}{X_{1c}^2+X_{1s}^2}X_{2c}+\\
        +\frac{X_{1s}Y_{1s}-X_{1c}Y_{1c}}{X_{1c}^2+X_{1s}^2}X_{20},
    \end{multline}
    \begin{multline}
        Y_{2c}=\frac{2X_{1c}X_{1s}}{X_{1c}^2+X_{1s}^2}Y_{20}+\frac{X_{1c}Y_{1c}+X_{1s}Y_{1s}}{X_{1c}^2+X_{1s}^2}X_{2s}+\frac{X_{1s}Y_{1c}-X_{1c}Y_{1s}}{X_{1c}^2+X_{1s}^2}X_{2c}-\\
        -\frac{X_{1s}Y_{1c}+X_{1c}Y_{1s}}{X_{1c}^2+X_{1s}^2}X_{20},
    \end{multline}
\end{subequations}
which are thus explicitly related to the second-order quantities $X_2$ and $Y_{20}$. This circumstance is reminiscent of the situation in a stellarator (Eqs.~(A32)-(A33) in \cite{landreman2019}). In the interest of brevity, we shall not proceed to find expressions for these coefficients. Full expressions may be found in the Zenodo repository associated with this paper.
\par
The remaining components of Eq.~(\ref{eqn:vacuum-mirror-cocon-inv}) can also be combined to yield expressions for $\Phi_2$ (i.e., three equations) and a consistency condition on first-order quantities. The former may be written as follows, using Eq.~(\ref{eqn:mirror-expansion-flux}),
\begin{subequations}
    \begin{align}
        \Phi_{20}=&\frac{B_0}{4}\frac{(X_{1c}^2+X_{1s}^2)(Y_{1s}X_{1c}'+Y_{1c}X_{1s}')+(Y_{1c}^2+Y_{1s}^2)(X_{1s}Y_{1c}'+X_{1c}Y_{1s}')}{X_{1s}Y_{1c}+X_{1c}Y_{1s}}, \\
        \Phi_{2c}=&\frac{B_0}{4}\frac{(X_{1c}^2-X_{1s}^2)(Y_{1s}X_{1c}'+Y_{1c}X_{1s}')+(Y_{1c}^2-Y_{1s}^2)(X_{1s}Y_{1c}'+X_{1c}Y_{1s}')}{X_{1s}Y_{1c}+X_{1c}Y_{1s}}, \\
        \Phi_{2s}=&\frac{B_0}{4}\frac{X_{1c}Y_{1c}[(X_{1s}^2)'+(Y_{1s}^2)'])+X_{1s}Y_{1s}[(X_{1c}^2)'+(Y_{1c}^2)'])}{X_{1s}Y_{1c}+X_{1c}Y_{1s}}+\frac{\Bar{B}}{2}\frac{X_{1c}X_{1s}'-X_{1s}X_{1c}'}{X_{1s}Y_{1c}+X_{1c}Y_{1s}},
    \end{align} \label{eqn:Phi_mirror_2nd_order}
\end{subequations}
and the latter as
\begin{equation}
    (X_{1c}^2-X_{1s}^2)(X_{1s}X_{1c}'-X_{1c}X_{1s}'+Y_{1s}Y_{1c}'-Y_{1c}Y_{1s}') =0.
\end{equation}
Because we have adopted an approach in which $X_{1s}$ and $X_{1c}$ are inputs to the near-axis construction, the second bracket must vanish. This is an ordinary differential equation for $Y_1$, which together with Eq.~(\ref{eqn:mirror-expansion-flux}) give both $Y_{1s}$ and $Y_{1c}$ in terms of $X_1$ and $B_0$. Introducing the definition $Y_{1c}=Y_{1s}\sigma$, this equation can be cast, with Eq.~(\ref{eqn:mirror-expansion-flux}), into the form of a Riccati equation,
\begin{equation}
    \sigma'=-(X_{1c}-X_{1s}\sigma)^2\left(\frac{B_0}{\bar{B}}\right)^2\left(X_{1s}X_{1c}'-X_{1c}X_{1s}'\right),
\end{equation}
while Eq.~(\ref{eqn:mirror-expansion-flux}) becomes
\begin{equation}
    Y_{1s}=\frac{\bar{B}}{B_0(X_{1c}-\sigma X_{1s})}.
\end{equation}

\subsubsection{Second order (\ref{eqn:jac-equation=-mirror})}
Having closed forms for the various terms in $\Phi_2$, we are in a position to evaluate the magnetic field $B_2$ by considering the $O(r^2)$ expansion of Eq.~(\ref{eqn:jac-equation=-mirror}). From each of its harmonics the following relations can be read off
\begin{subequations}
    \begin{align}
        \frac{B_{20}}{B_0}=&\frac{\Phi_{20}'}{B_0}-\frac{1}{4}\left[(X_{1c}')^2+(X_{1s}')^2+(Y_{1c}')^2+(Y_{1s}')^2\right], \\
        \frac{B_{2c}}{B_0}=&\frac{\Phi_{2c}'}{B_0}-\frac{1}{4}\left[(X_{1c}')^2-(X_{1s}')^2+(Y_{1c}')^2-(Y_{1s}')^2\right],\\
        \frac{B_{2s}}{B_0}=&\frac{\Phi_{2s}'}{B_0}-\frac{1}{2}\left(X_{1c}'X_{1s}' + Y_{1c}'Y_{1s}'\right),
    \end{align} \label{eqn:mirror-nae-B2}
\end{subequations}
which uniquely describe the magnetic field magnitude to second order in terms of first-order shaping. Note that the quadrupole nature of $|\mathbf{B}|$ arises naturally through the near-axis construction.

\subsection{Magnetic field at the bottom of the well}
We now have all the tools we need to understand the behaviour of the field at the bottom of the magnetic well, which we define to be at $z=0$. We choose to align our coordinate system with the leading elliptical shape at $z=0$ and define the poloidal label $\alpha$ in such a way that $\sigma(0)=0$ and $X_{1s}(0)=0$, without any loss of generality.\footnote{We could have kept $\sigma(0)$ non-zero in order to mirror the stellarator case, but in the interest of reducing the algebra we choose it to vanish. The reason why in the stellarator case this choice is not general is that the basis used is the Frenet-Serret frame, which we are not free to redefine.}  By definition, $B_0'(0)=0$ and we shall choose our reference field as $B_0(0)=\bar{B}$. 
\par
With these definitions and choices, we may then evaluate the magnetic field at the bottom of the magnetic well for a general straight magnetic mirror. The result is
\begin{subequations}
    \begin{gather}
        B_{20}(z=0)=\frac{B_0}{4}\left\{-\frac{B_0''}{B_0 X_{1c}^2}+2\left[\left(\frac{X_{1c}'}{X_{1c}^2}\right)^2+(X_{1s}')^2\right]+X_{1c}''X_{1c}\left(1-\frac{1}{X_{1c}^4}\right)\right\}, \\
        B_{2c}(z=0)=\frac{B_0}{4}\left\{\frac{B_0''}{B_0 X_{1c}^2}-2\left[\left(\frac{X_{1c}'}{X_{1c}^2}\right)^2+(X_{1s}')^2\right]+X_{1c}''X_{1c}\left(1+\frac{1}{X_{1c}^4}\right)\right\}, \\
        B_{2s}(z=0)=\frac{B_0}{2}X_{1c}X_{1s}'',
    \end{gather} \label{eqn:B2_bottom_mirror}
\end{subequations}
where all quantities are evaluated at $z=0$.

\subsection{Maximum-$J$ in a magnetic mirror}
The analysis of the maximum-$J$ condition in a mirror closely follows that in a QI stellarator and proceeds from the general expression for the precession frequency (\ref{omegaa}). In the context of a straight mirror, relaxing the conditions of omnigeneity and stellarator symmetry, but instead noting that there is no first order variation in the field magnitude, $B_1$, we find in vacuum
 \begin{equation}
        \omega_\alpha = \frac{2mv^2}{q \bar{B}}\frac{\int_{z_{L}}^{z_{R}}\frac{1-\lambda B_0/2}{B_0\sqrt{1-\lambda B_0}}B_2(z,\alpha)\mathrm{d}z}{\int_{z_{L}}^{z_{R}}\frac{\mathrm{d}z}{\sqrt{1-\lambda B_0}}}, \label{eqn:nae_w_alpha_mirror}
    \end{equation}
where $B_2=B_{20}+B_{2c}\cos2\alpha+B_{2s}\sin2\alpha$, $z_{L,R}$ are the bouncing points for given $\lambda$ and the magnetic field functions are given in Eqs.~(\ref{eqn:mirror-nae-B2}). We may normalise $\omega_\alpha$ following the prescription in Eq.~(\ref{omegaa-hat}), for which we need to define an edge flux $\psi_a$. Following the same procedure as for the near-axis stellarator, defining the aspect ratio as $A=L/a$, where $L$ is the length of the mirror, 
\begin{equation}
    \psi_a=\frac{1}{2}\frac{L^3}{A^2}\left(\int\frac{\mathrm{d}z}{B_0}\right)^{-1}.
\end{equation}
\par
At the bottom of the well, the expression for $\omega_\alpha$ for deeply trapped particles reduces to a form analogous to Eq.~(\ref{eqn:deep-QI-wa-nae}),
\begin{equation} 
    \omega_\alpha \approx \frac{mv^2}{qB_0^2} \left( B_{20} + B_{2c}\cos2\alpha+B_{2s}\sin2\alpha \right). \label{eqn:deep-mirror-wa-nae}
\end{equation}
For maximum-$J$ behaviour, i.e. $q\omega_\alpha>0$ for all $\alpha$, it is thus necessary that
\begin{equation}
    B_{20}^2>B_{2c}^2+B_{2s}^2,
\end{equation}
in addition to $B_{20}>0$. Because the mirror has no first-order variation of $|\mathbf{B}|$, the maximum-$J$ and minimum-$B$ properties are equivalent, i.e. both require $\partial_\psi B>0$ to leading order in the expansion in $r$. This minimum-$B$ property is well known to endow the magnetic mirror with MHD stability at sufficiently low $\beta$ \citep{berkowitz1958mhd,Taylor_1963,furth1964closed}.
\par
Using the expressions for $B_2$ at the bottom of the mirror, Eqs.~(\ref{eqn:B2_bottom_mirror}), one may show that the maximum-$J$ condition is equivalent to 
\begin{subequations}
    \begin{gather}
        X_{1c}''X_{1c}>0, \label{eqn:min-B-mirr-i} \\
        \left(\frac{X_{1c}'}{X_{1c}^2}\right)^2+(X_{1s}')^2-\frac{B_0''}{2B_0X_{1c}^2}>\frac{X_{1c}X_{1c}''}{2}\left[\frac{1}{X_{1c}^4}+\left(\frac{X_{1s}''}{X_{1c}''}\right)^2\right]. \label{eqn:min-B-mirr-ii}
    \end{gather}
\end{subequations}
These two conditions may qualitatively be interpreted as
\begin{subequations}
    \begin{gather*}
        \text{Elongation~increase}\neq 0, \\
        \mathrm{Twist}^2~-~\frac{\mathrm{Mirror~ratio}}{\mathrm{Well~width}^2} > \mathrm{Elongation~increase}.
    \end{gather*}
\end{subequations}
Eq.~(\ref{eqn:min-B-mirr-i}) is a statement about the elongation of flux surfaces having to grow away from the middle of the mirror, where $X_{1c}^2$ is the elongation of the elliptical cross-section in the $x$-direction. We are free to take $X_{1c}>1$ to make this direction coincide with the major axis of the ellipse. Then $X_{1c}''>0$ can be interpreted as a condition for growing elongation away from the minimum (in both directions). This property indeed constitutes a typical feature of optimised magnetic mirrors (see Fig.~\ref{fig:opt-mirror-3d}).
\par
Equation (\ref{eqn:min-B-mirr-ii}) is more interesting than Eq.~(\ref{eqn:min-B-mirr-i}), as it involves more aspects of the field. Maximum-$J$ behaviour can only be attained if there is a sufficiently large `twist' of field lines across the bottom of the magnetic well. By `twist' we refer to the contribution from $X_{1c}'$ and $X_{1s}'$, which describes a left-right symmetry breaking of the field. This `local shearing' of field lines and lack of symmetry about the minimum of the configuration is common in mirror designs. The twist of the field must be strong enough to overcome the effect of elongation and the natural tendency for a negative radial magnetic field derivative, indicated by the $B_0''$ term.
\par
All in all, strong twist and large elongation are helpful for maximum-$J$ behaviour, but they must be balanced in a careful way. 
Making the minimum flat (reducing $B_0''$) is also helpful. These features are visible in typically shaped mirrors such as the example shown in Fig.~\ref{fig:opt-mirror-3d}.

\subsection{Omnigeneity in a magnetic mirror}
Magnetic mirrors can be made to be omnigenous by careful tailoring of the quadrupole magnetic field \citep{Hall,Catto1981}. What this means in an open ended device is that the surfaces of constant $\psi$ correspond to precession surfaces for all trapped species in the sense introduced by \cite{Hall}.
In order to characterise this property, we measure of the `radial' drift (normal to constant $\psi$ surfaces) by the quantity $Y_\mathrm{omn}=\nabla\Phi\times\nabla B\cdot\nabla\psi/\mathbf{B}\cdot\nabla B$, which plays a key role in the analysis of stellarator omnigeneity and in our coordinates $(\psi,\alpha,z)$ can be written as
\begin{equation}
    Y_\mathrm{omn}=\partial_\alpha\Phi - \partial_z\Phi\frac{\partial_\alpha B}{\partial_z B}.
\end{equation}
We follow the approach of \cite{rodriguez2023higher} to obtain a near-axis expansion of the condition for omnigeneity, which requires that $Y_\mathrm{omn}$ is equal and opposite at bounce points.
\par
Expanding this condition in powers of $r$, we obtain $Y_\mathrm{omn}^{(1)}=0$ to first order by virtue of the straight axis: there is no leading curvature drift. It is only at the next order that there is a non-zero radial drift, so we may write $Y_\mathrm{omn}^{(2)}=Y_\mathrm{omn,c}^{(2)}\cos 2\alpha+Y_\mathrm{omn,s}^{(2)}\sin 2\alpha$. The expansion gives
\begin{subequations}
\begin{align}
    Y_\mathrm{omn,c}^{(2)}&=2\left(\Phi_{2s}-\frac{B_0}{B_0'}B_{2s}\right)=-2\frac{B_0^2}{B_0'}\left[\left(\frac{\Phi_{2s}}{B_0}\right)'+V_2\right], \\
    Y_\mathrm{omn,s}^{(2)}&=-2\left(\Phi_{2c}-\frac{B_0}{B_0'}B_{2c}\right)=2\frac{B_0^2}{B_0'}\left[\left(\frac{\Phi_{2c}}{B_0}\right)'+V_1\right], \label{eqn:Yomn-mirror-ii}
\end{align}\label{eqn:Yomn-mirror}
\end{subequations}
where,
\begin{subequations}
    \begin{gather}
        V_1=\frac{1}{4}\left[(Y_{1s}')^2+(X_{1s}')^2-(Y_{1c}')^2-(X_{1c}')^2\right], \\
        V_2=-\frac{1}{2}\left(X_{1c}'X_{1s}'+Y_{1c}'Y_{1s}'\right), \\
    \end{gather} \label{eqn:Yomn2}
\end{subequations}
and we have made use of Eq.~(\ref{eqn:mirror-nae-B2}). For omnigeneity, the expressions for $Y_\mathrm{omn,c}^{(2)}$ and $Y_\mathrm{omn,s}^{(2)}$ must be even in $z$ (provided $B_0$ is symmetric about its minimum). This imposes constraints on the choice of $X_{1c}$ and $X_{1s}$.
\par
It is natural to ask whether this condition is at all compatible with the maximum-$J$ requirements devived above. In other words, is it possible to design an omnigenous, maximum-$J$ magnetic mirror? Let us consider omnigeneity at the bottom of the mirror, in which the condition reduces to that of locally vanishing radial drift. Evaluating the square brackets in Eqs.~(\ref{eqn:Yomn-mirror}) at the minimum, and requiring them to vanish (assuming a first order zero of $B_0''\neq0$),
\begin{subequations}
    \begin{gather}
        X_{1s}''=0, \\
        \left(\frac{X_{1c}'}{X_{1c}^2}\right)^2+(X_{1s}')^2=\frac{B_0''}{2B_0X_{1c}^2}+\frac{X_{1c}X_{1c}''}{2}\left(1+\frac{1}{X_{1c}^4}\right), \label{eqn:mirror-min-omn-ii}
    \end{gather}
\end{subequations}
which, incidentally, implies $B_{20}/B_0=X_{1c}''X_{1c}/2$ and $B_{2c}=0=B_{2s}$ at the minimum. If we choose $X_{1c}''>0$ to comply by the first of the two maximum-$J$ conditions, Eq.~(\ref{eqn:min-B-mirr-i}), the second maximum-$J$ condition, Eq.~(\ref{eqn:min-B-mirr-ii}), is satisfied by satisfying the condition of omnigeneity, Eq.~(\ref{eqn:mirror-min-omn-ii}). This suggests that it is possible to construct a mirror that is omnigeneous and satisfies the maximum-$J$ condition simultaneously, at least at the minimum. Indeed, \cite{Catto1981} have shown that it is possible in a finite region around the minimum. 
\par

\subsection{Example of approximately omnigeneneous and maximum-$J$ mirror}
\begin{figure}
    \centering
    \includegraphics[width=0.7\textwidth]{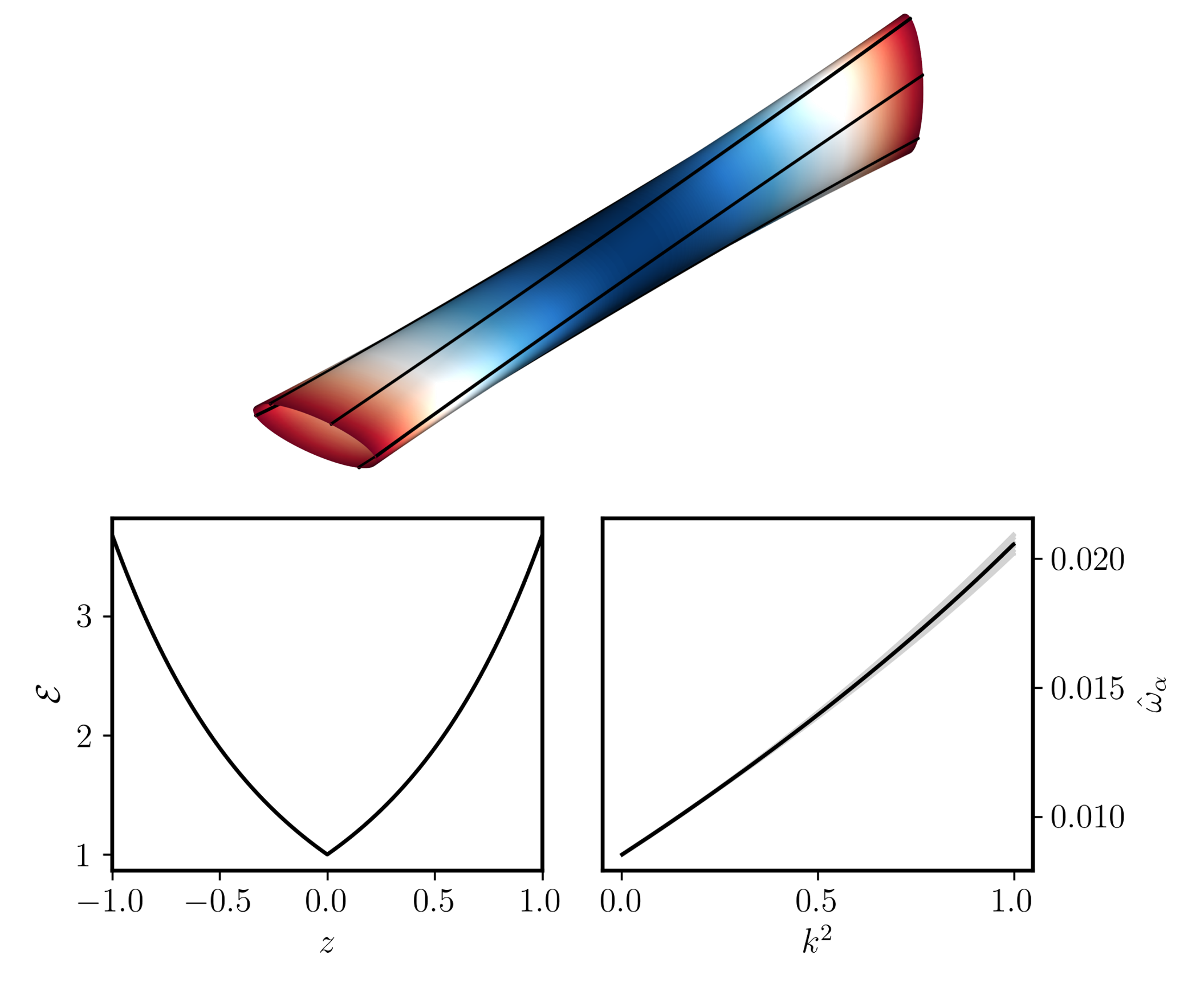}
    \caption{\textbf{Example of an optimised `symmetric' magnetic mirror.} The figure shows a 3D rendition of a near-axis magnetic mirror optimised for omnigeneity and the maximum-$J$ property. The bottom plots illustrate the elongation of flux surfaces, $\mathcal{E}$, and the normalised precession frequency (with $\hat{\omega}_\alpha>0$ indicating maximum-$J$ behaviour) as a function of $z$, where the black line represents the average over the field-line label $\alpha$. The mirror was constructed assuming symmetry of $X_{1c}$ and $X_{1s}$, and employs a simple quadratic magnetic field.}
    \label{fig:opt-mirror-3d}
\end{figure}

To venture beyond the bottom of the well, we present a numerical example in which the maximum-$J$ and omnigeneity conditions are both satisfied over a non-zero interval in $z$. To construct such an example, we choose for simplicity a symmetric solution with well defined parity (so that $X_{1s}$ and $X_{1c}$ are odd and even respectively). This guarantees the correct behaviour of Eq.~(\ref{eqn:Yomn-mirror-ii}), but leaves us free to find inputs such that $V_1+(\Phi_{2c}/B_0)'=0$. This task can be formulated as an optimisation problem in which $X_{1s}$ and $X_{1c}$ are the degrees of freedom (fixing $B_0$). We initialise the search with a configuration that satisfies the omnigeneity requirement at the bottom of the well (as we know in closed form what this choice should be). The result of the optimisation (with 6 scalar degrees of freedome) is presented in Fig.~\ref{fig:opt-mirror-3d}.
\par

\section{Details of near-axis optimisation for a vacuum QI field with the maximum-$J$ property} \label{sec:app_opt_nae}
In this Appendix, we document details of the optimisation approach to the near-axis field presented in Figure~\ref{fig:opt_nae_maxj} of the main text. A more in-depth discussion will be presented elsewhere.
\par
The main goal of our optimisation effort is to construct a proof-of-concept near-axis field that exhibits the maximum-$J$ property in the omnigenous portion of the trapped-particle population. In that regard, we do not optimise for the particles in the so-called \textit{buffer-region} \citep{plunk2019direct}, the part of the domain where omnigeneity is violated in order to enforce periodicity of the magnetic field. In general, narrow buffer regions lead to large shaping gradients, and these regions are therefore often made relatively wide although omnigeneity then suffers. As far as the maximum-$J$ property is concerned, this will increase the number of particles with large ($\sim 1/r$) detrimental precession. We shall however not be concerned with this detail but argue that we could, in principle, shrink the buffer region to become arbitrarily small.
\par
Focusing, then, on the other contributions to the precession, we  need to introduce a measure for maximum-$J$ behaviour that serves as a target function in the numerical optimisation. In order to exclude contributions from the buffer regions proportional to $1/r$, we evaluate the precession frequency as the sum of Eqs.~(\ref{eqn:nae_w_alpha}) and (\ref{eqn:2nd-order-deviation-wa}), and construct the target function as the sum
 $$ g_{\omega_\alpha} = \sum_{k}\hat{\omega}_\alpha^2 (k), $$
 taken over values of $k$ such that $\hat \omega_\alpha < 0$. The required integrals are evaluated using the help of the \texttt{BAD} library in \texttt{python} \citep{mackenbach2023bad}. 
\par
A second ingredient in the optimisation is the condition of omnigeneity (outside the buffer regions). Given first-order inputs, second order choices necessary to enforce the correct form for $B_{2c}$. Accordingly, we construct $X_{2c}$ and $X_{2s}$ following Eqs.~(\ref{eqn:X2c}) and the equivalent for $X_{2s}$, assuming that $B_{2c}$ has the correct form for QI. Writing $X_{2c}=\Tilde{X}_{2c}c_{2\alpha_1}+\Tilde{X}_{2s}s_{2\alpha_1}$ and $X_{2s}=\Tilde{X}_{2c}s_{2\alpha_1}-\Tilde{X}_{2s}c_{2\alpha_1}$, we find
\begin{subequations}
    \begin{gather}
        \tilde{X}_{2c}=\frac{1}{\kappa B_0}\left[\hat{\mathcal{T}}_c c_{2\alpha_1}+\hat{\mathcal{T}}_s s_{2\alpha_1}+\frac{B_0^2}{4}\left(\frac{d^2}{B_0'}\right)'\right], \label{eqn:X2ctilde}\\
        \tilde{X}_{2s}=\frac{1}{\kappa B_0}\left[\hat{\mathcal{T}}_c s_{2\alpha_1}-\hat{\mathcal{T}}_s c_{2\alpha_1}-B_{2s}^\mathrm{QI}\right]. \label{eqn:X2stilde}
    \end{gather}
\end{subequations}
where $c_{2\alpha_1}=\cos2\alpha_1$ and $s_{2\alpha_1}=\sin2\alpha_1$, and
\begin{subequations}
    \begin{gather}
        \hat{\mathcal{T}}_c=\frac{B_0}{l'}\left[Z_{2c}'+2\iota_0 Z_{2s}+\frac{q_c^2-q_s^2+r_c^2-r_s^2}{4l'}\right], \\
        \hat{\mathcal{T}}_s=\frac{B_0}{l'}\left[Z_{2s}'-2\iota_0 Z_{2c}+\frac{q_cq_s+r_cr_s}{2l'}\right],
    \end{gather}
\end{subequations}
and $\alpha_1=\nu-\iota\varphi$ in the notation of Appendix~\ref{app:precession}, following \cite{garrenboozer1991a}. This construction guarantees QI behaviour at second order for any odd $B_{2s}^\mathrm{QI}$, as can be checked by substitution. The only exceptions are points where the curvature vanishes, where we learnt in Appendix~\ref{sec:appB20Deep}, Eq.~(\ref{eqn:omnnn-constr-min}), that the QI condition reduces to a condition on lower-order coefficients. An indication of the latter is that the construction will generally diverge wherever $\kappa=0$, from which more general expressions than those of Eq.~(\ref{eqn:omnnn-constr-min}) can be obtained which do not assume QI at first order (this is important for the tops of the well in practice). Thus, one must add to the optimisation a measure of the deviation from QI at the points of minimum and maximum field strength, which the optimiser should attempt to reduce. We denote the sum of this `residual' at all extrema along the field by $\check{g}_\mathrm{QI}$. To simplify the available choices, we further take $p_2=0$ and $\tilde{X}_{2s}=0$, which is not the most general choice, and most likely not even the best, but suffices for our purposes. Although this approach enforces QI at second order exactly, it does not appear to be the best one in practice, as it leads to highly shaped configurations, and unnecessarily enforces QI at second order in regions where it is already violated at first order. However, it suffices for our purposes of proof-of-principle construction, and we leave further refinements for future work.
\par
The target function to be numerically is thus $g=g_{\omega_\alpha}+\check{g}_\mathrm{QI}$, and we proceed to discuss the degrees of freedom over which the optimisation is to be carried out. At first order, there are many free functions, some of which will be fixed for simplicity. The magnetic field strength is taken to vary along the axis as $B_0=1+0.16\cos\varphi$, but the shape of the magnetic axis and the elongation measure $\bar{d}$ are allowed to change. For the former, we vary three even Fourier harmonics (three for each $Z$ and $R$ describing the axis in cylindrical coordinates), a fourth one being chosen consistently to make the curvature vanish at $\phi=0,~\pi$ \citep{rodriguez2022phases,camacho-mata-2022}. This restricts us to first order zeroes of curvature \citep{rodriguez2022phases,mata2023helicity}. The function $\bar{d}$ is allowed to have 7 degrees of freedom through collocation points of a symmetric spline, symmetry being necessary to preserve stellarator symmetry. 
\par
With the optimisation procedure defined this way, we finally need to specify a starting point, which we take to be a generic QI axis with $N=1$ and $\bar{d}=1$. Other details, such as the buffer region size etc. can be found in the supplementary material, in which all the numerical tools are provided. The optimisation is performed using the near-axis code \texttt{pyQIC} \citep{pyQIC} with appropriate upgrades (use of splines, second order choice, etc.), and the optimisation libraries of \texttt{scipy} (the algorithms Nelder-Mead and BFGS are used, the latter as a refinement of the former). Broadly speaking, the optimisation proceeds in two stages: (i) Nelder-Mead optimisation for QI and maximum-$J$ behaviour at the minima and maxima of $B_0$; (ii) full optimisation (on a few field lines) of the function $g$. Step (i) has the benefit of only requiring a first-order near-axis solution, thus being quite fast, while (ii) is slower as it needs to compute $\hat{\omega}_\alpha$. Additional refinements are of course possible.
\par
The results of the optimisation are presented in Figure~\ref{fig:opt_nae_maxj}, and all the pertinent files and scripts are included in the Zenodo repository associated with this paper.

\bibliographystyle{jpp}

\bibliography{jpp-instructions}

\end{document}